\documentclass[fleqn,10pt]{wlscirep}
\usepackage[utf8]{inputenc}
\usepackage[T1]{fontenc}
\usepackage[square,sort,comma,numbers]{natbib}

\usepackage{url,booktabs}
\title{COVID-19 -- A simple statistical model for predicting intensive care unit load in early phases of the disease}

\author[a]{Matthias Ritter}
\author[b]{Derek V.M. Ott}
\author[c]{Friedemann Paul}
\author[d]{John-Dylan Haynes}
\author[d,e,*]{Kerstin Ritter}

\affil[a]{Humboldt-Universität zu Berlin, Faculty of Life Sciences, Unter den Linden 6, 10099 Berlin, Germany}
\affil[b]{Neurology Clinic with
Stroke Unit and Early Rehabilitation, Unfallkrankenhaus Berlin, 12683 Berlin, Germany}
\affil[c]{Charité - Universitätsmedizin Berlin, corporate member of Freie Universität Berlin, Humboldt-Universität zu Berlin, and Berlin Institute of Health (BIH); Department of Neurology, Experimental and Clinical Research Center and Max Delbrueck Center for Molecular Medicine, Charitéplatz 1, 10117 Berlin, Germany}
\affil[d]{Charité - Universitätsmedizin Berlin, corporate member of Freie Universität Berlin, Humboldt-Universität zu Berlin, and Berlin Institute of Health (BIH); Berlin Center for Advanced Neuroimaging, Bernstein Center for Computational Neuroscience, Charitéplatz 1, 10117 Berlin, Germany}
\affil[e]{Charité - Universitätsmedizin Berlin, corporate member of Freie Universität Berlin, Humboldt-Universität zu Berlin, and Berlin Institute of Health (BIH); Department of Psychiatry and Psychotherapy, Charitéplatz 1, 10117 Berlin, Germany}

\affil[*]{kerstin.ritter@charite.de}

\begin{abstract}
One major bottleneck in the ongoing COVID-19 pandemic is the limited number of critical care beds. Due to the dynamic development of infections and the time lag between when patients are infected and when a proportion of them enters an intensive care unit (ICU), the need for future intensive care can easily be underestimated. To infer future ICU load from reported infections, we suggest a simple statistical model that (1) accounts for time lags and (2) allows for making predictions depending on different future growth of infections. We have evaluated our model for three regions, namely Berlin (Germany), Lombardy (Italy), and Madrid (Spain). Before extensive containment measures made an impact, we first estimate the region-specific model parameters. Whereas for Berlin, an ICU rate of 6\%, a time lag of 6 days, and an average stay of 12 days in ICU provide the best fit of the data, for Lombardy and Madrid the ICU rate was higher (18\% and 15\%) and the time lag (0 and 3 days) and the average stay (4 and 8 days) in ICU shorter. The region-specific models are then used to predict future ICU load assuming either a continued exponential phase with varying growth rates (0–-15\%) or linear growth. Thus, the model can help to predict a potential exceedance of ICU capacity. Although our predictions are based on small data sets and disregard non-stationary dynamics, our model is simple, robust, and can be used in early phases of the disease when data are scarce.
\end{abstract}
\begin{document}

\flushbottom
\maketitle
%
%
\thispagestyle{empty}


\section*{Introduction}

The number of reported COVID-19 cases worldwide is steadily increasing and has reached 14.5 million on July 19, 2020.\footnote{The numbers of daily infections are reported by the Corona Resource Center of the John Hopkins University: \url{https://coronavirus.jhu.edu/map.html}} On March 26, 2020, the U.S. became the country with the largest number of officially reported infections (83.800 people). Until July 19, 2020, this number increased to 3.7 million people. Even though these numbers strongly depend on the number of conducted tests, they demonstrate the huge spread and severity of the SARS-CoV-2 pandemic.  

It lies in the nature of exponential growth that it starts slowly and bears the risk that the future development is underestimated, as shown in multiple psychological studies \citep{Wagenaar1975}. In the SARS-CoV-2 pandemic, this leads to the risk of underestimating case fatality rates \citep{Baud2020,Lipsitch2020} but also the risk that the health system might be overburdened due to a too high number of patients in intensive care units (ICUs; \citep{Grasselli2020}). Although it is well-known that a certain number of COVID-19 patients needs intensive care, especially elderly people and people with pre-existing conditions \citep{Verity2020}, the exponential dynamics of infections along with the time lag between the number of reported infections and the number of ICU patients can lead to the false impression that the amount of ICU patients will be unproblematic \citep{Baud2020,Ferriss2020}. Due to the lag and ICU durations of one to several weeks, even when the exponential growth of infections is stopped, it takes a while until the pressure on ICUs is reduced \citep{Bhatraju2020,Clukey2020,Manca2020}.  

In this early phase of the SARS-CoV-2 pandemic, in which epidemiological data is non-stationary, heterogeneous, and regionally specific, the prediction of future ICU load (as well as mortality) is a highly challenging task \citep{Deasy2020,Ferguson2020}. In addition to certain assumptions (e.g., ICU rate or length of ICU stay) that have to be estimated from data or pre-specified, the dynamics of exponential growth are unforeseeable in times where containment measures ranging from closure of schools to home office or curfew have been applied to alleviate exponential growth \citep{Ferguson2020}. In most forecast models, however, the exponential growth is assumed to continue with the same rate over the prediction horizon (mostly 14 days) and is fitted either to the daily incidence of COVID-19 patients or ICU patients \citep{Deasy2020,Grasselli2020,Remuzzi2020}.  

In this study, we suggest a simple and transparent statistical model that is able to account for (1) the time lag between reported infection and ICU admission and (2) different future growth (linear or exponential). This allows us in particular to predict the time point when the ICU load exceeds a given capacity. We address the following research questions: Can the future number of ICU patients be predicted from the number of reported infections? And how does the growth rate influence the time when ICU capacity is expected to be overburdened? 

In our empirical examples, we apply the model to three public data sets, namely Berlin (Germany), Lombardy (Italy), and Madrid (Spain). Given the total number of COVID-19 patients and the number of COVID-19 patients in ICUs, we first estimate the parameters of the model (i.e., ICU rate, length of stay in ICU, and time lag between the positive testing and the ICU admission). We restricted here the data to the time period before extensive containment measures made an impact.
And second, we predict for a time horizon of two months the number of COVID-19 patients who need intensive care when assuming linear or exponential growth of the number of reported infections. By incorporating the dependency between the total number of COVID-19 patients and COVID-19 patients in ICUs as well as holding growth rates flexible, our model extends previous models relying on fitting linear or exponential growth to initial ICU case data \citep{Grasselli2020,Remuzzi2020}. The predictions of our model in combination with epidemiological estimates regarding disease dynamics can be used in turn to predict future ICU load and to potentially predict when the ICU load will exceed a given capacity.

\section*{Method}
\label{S:2}

In this section, we introduce the theoretical model for predicting the number of ICU patients based on the number of reported infections. Denote $PT_{t}$ the total number of positively tested people until day $t$ in a particular region.
$\Delta PT_{t}=PT_{t}-PT_{t-1}$
is then the number of newly positively tested people only on day
$t$. A certain share $\alpha_{l}$ of the newly positively tested people needs intensive care $l$ days later with lag $l, l=1,...,L$, and some maximum lag $L$ denoting the maximal duration between a positive test and ICU admission. Moreover, patients remain in intensive care for a longer time, which means that ICU admissions of the previous days also have to be considered. In early stages of a disease the exact distribution of the lengths of ICU stays might not be available and for simplification, the duration can be modelled as constant for all patients. We denote $K$
the average number of days patients remain in an ICU.  

The number of ICU patients at time
$t$ can then be predicted in dependence of $K$
and a vector $\alpha_{l}$ with
$l=1,\dots,L$, which contains the probabilities that positively tested persons have transited to ICU after
$l$ days, in the following way: 

\begin{equation} 
\widehat{IC}_{t}(K, \alpha_{l})=\sum_{k=1}^{K}\sum_{l=1}^{L}\alpha_{l}\Delta PT_{t-l-k+1}
\end{equation}

In case only the overall share of reported infected people needing intensive care without differentiating between the lags, the so-called ICU rate $\alpha$ with $\alpha=\sum_{l=1}^{L}\alpha_{l}$, is available, this can be used in combination with a specific average lag
$l^{*}$. Eq. (1) then becomes: 
\begin{equation} 
\widehat{IC}_{t}(K, \alpha, l^{*})=\sum_{k=1}^{K}\alpha\Delta PT_{t-l^{*}-k+1}
\end{equation}

Since Eq. (2) corresponds to the data situation in our empirical application, the subsequent considerations will be based on this equation. They can, however, analogously be applied to Eq. (1). 

To evaluate the performance of the models and to derive the parameters ICU rate $\alpha$, the length of ICU stay $K$ and the lag length $l^{*}$ that best explain the data, the squared correlation coefficient and the root mean squared prediction error (RMSE) for each model can be calculated by comparing the predicted value
$\widehat{IC}_{t}(K, \alpha, l^{*})$ to the observed value ${IC}_{t}$.
The RMSE is defined as 
\begin{equation}
    \text{RMSE}(K, \alpha, l^{*})=\sqrt{\sum_{t=1}^{T}(IC_{t}-\widehat{IC}_{t}(K, \alpha, l^{*}))^2}
\end{equation}
with $T$ denoting the number of days with observations available. The model with the lowest RMSE is supposed to fit best to the observed data.  

After the parameters $K$, $a$, and $l^{*}$ are estimated (based on prior knowledge or data), they can be used to predict the future development by assuming either linear growth with slope $d$ for the number of infections, i.e., $PT_{t+1}=PT_t+d$, or exponential growth with rate $r$ for the number of infections, i.e.,
$PT_{t+1}=(1+r)PT_{t}$, which also holds for the daily changes, $\Delta PT_{t+1}=(1+r)\Delta PT_{t}$. The predicted number of ICU patients can then be compared with different levels of ICU capacities. Moreover, dates when a given capacity is expected to be exceeded can be calculated.

\section*{Empirical applications}
We now apply the above introduced theoretical model to data of the initial phase of the SARS-CoV-2 pandemic for Berlin (Germany), Lombardy (Italy), and Madrid (Spain). In a first step, we determine the parameters $K$, $a$, and $l^{*}$ that best explain the data in each region. 
We then use this model to predict the number of ICU patients for a fixed time horizon of two months assuming further linear or exponential growth. Please note that the model is sensitive to the estimated parameters and the underlying data. Since we are in a dynamic and non-stationary situation where the data and the development can change on a daily basis, we also demonstrate the sensitivity of the development regarding future growth and different time periods within the SARS-CoV-2 pandemic.

\subsection*{Data}
\label{sec:Data}
\subsubsection*{Berlin (Germany)}

\begin{figure}
\centering
\includegraphics[width=0.49\linewidth]{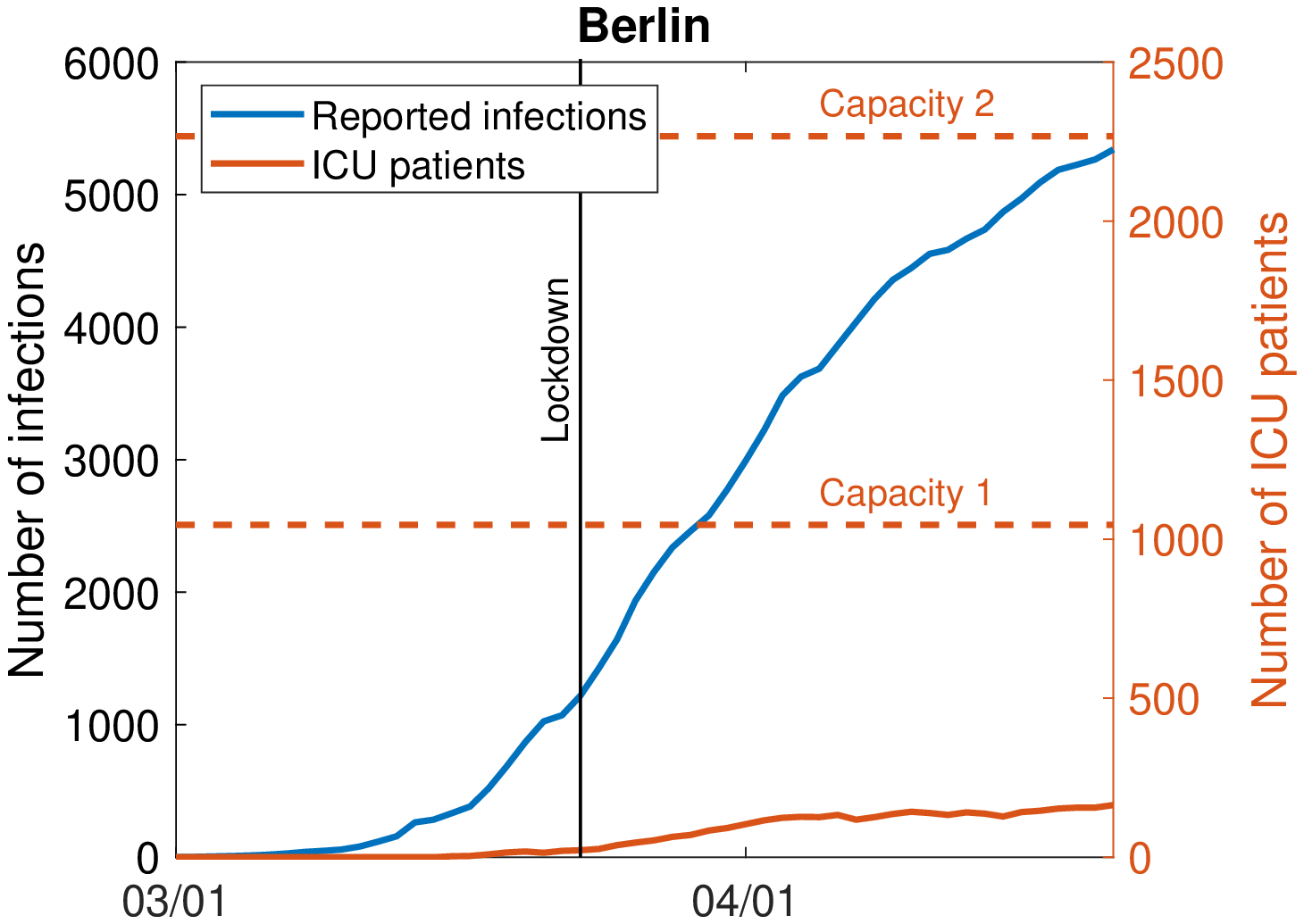}
\includegraphics[width=0.49\linewidth]{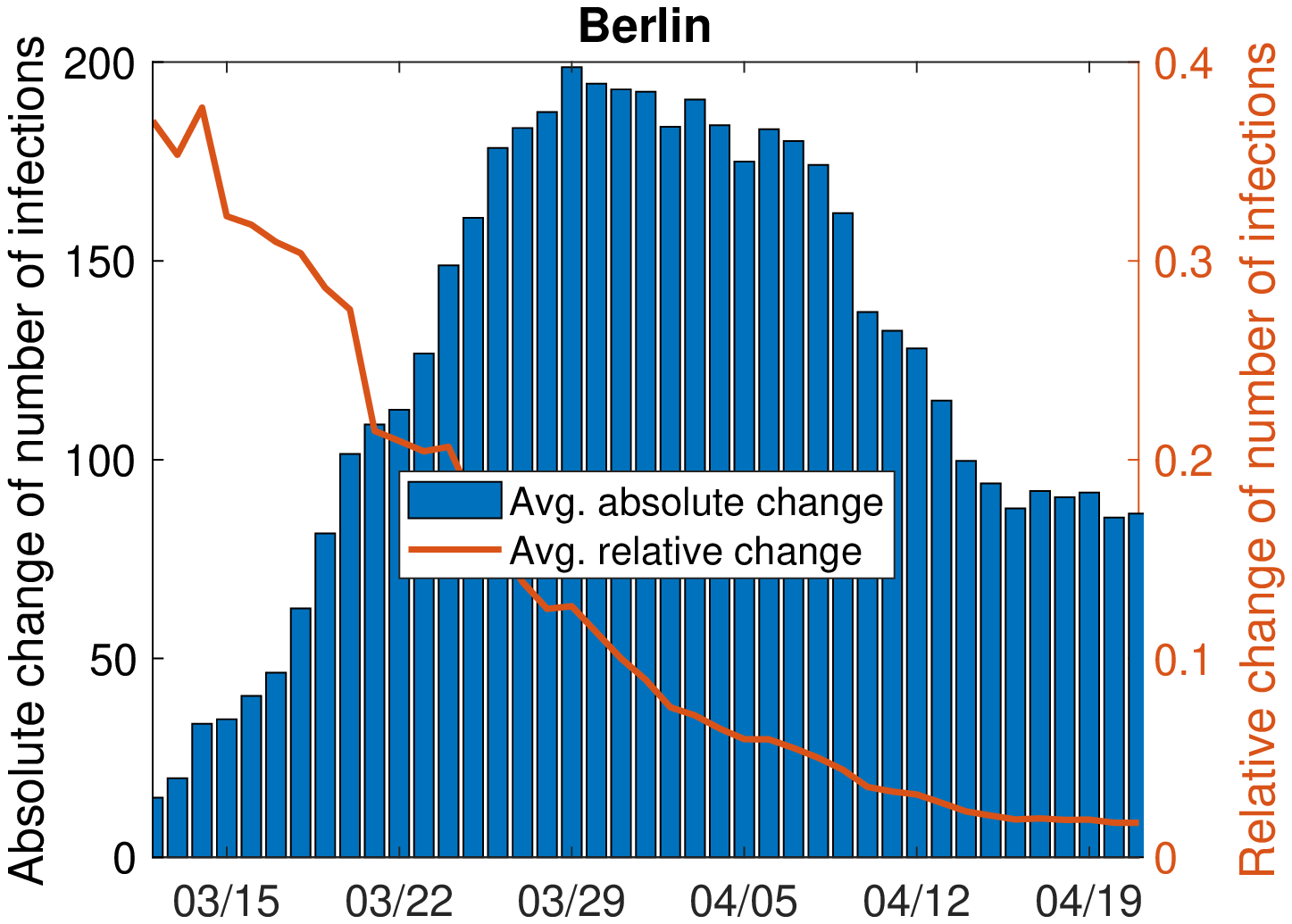}
\caption{Number of reported infections and intensive care patients (left) and absolute and relative daily change of the number of infections (right, averaged over the last 7 days, since 100 infections) in Berlin until April 21, 2020 (based on data from the \textit{Berlin Senate Department for Health, Nursing and Equal Opportunities})}
\label{Figure:Data_Berlin}
\end{figure}

Berlin is one of the 16 federal states of Germany and is densely populated with currently about 3.8 million inhabitants \citep{StatBerlin}. For each day between March 1 and April 21, we retrieved the number of reported COVID-19 patients and the number of COVID-19 patients in ICUs for Berlin from the \textit{Berlin Senate Department for Health, Nursing and Equal Opportunities} (Figure \ref{Figure:Data_Berlin}, left).\footnote{\url{https://www.berlin.de/sen/gpg/service/presse/2020/}} The total number of COVID-19 infections increased from one on March 1, 2020, to 5,341 on April 21, 2020. The first patients in intensive care (three) were reported on March 16, 2020, and this number increased to 164 on April 21, 2020.  
Figure \ref{Figure:Data_Berlin} (right) depicts the (absolute and relative) daily change in the number of reported infections over the previous 7 days since the first day more than 100 cases were reported (March 12, 2020, 118 cases). This averaging rules out weekend effects when some numbers are reported with delay.

On March 23, 2020, extensive containment measures including school closure and contact restriction were established for Berlin.\footnote{\url{https://www.berlin.de/sen/justiz/service/gesetze-und-verordnungen/2020/ausgabe-nr-12-vom-27-3-2020-s-217-224.pdf}} 
To make the three regions comparable and to ensure a greater homogeneity of the data, we restrict our analysis to the period before containment measures make an impact. For Berlin, we therefore include only data from March 1, 2020, to March 30, 2020. The end date corresponds to 7 days after the lockdown. 
The overall daily growth rate for this period (with a minimum of 100 cases) is 19\% ($\sqrt[18]{2{,}581/118}=1.19$), but it decreased to 11\% in the last week ($\sqrt[7]{2{,}581/1{,}219}=1.11$). Due to the containment measures, the average daily growth further decreased since then. 


The deployment of ICU beds is a highly dynamic process. According to the health senator of Berlin, Dilek Kalayci, the number of ICU beds in Berlin was 1,045 in March 2020 and has been planned to increase to 2,267 until the end of April \citep{Schroeter2020, SenGPG2020}.\footnote{For 2017, the Office for Statistics of Berlin-Brandenburg reported 1,450 ICU beds \citep{StatBund2018}. However, since this number is already 4 years old, we prefer to use the current estimates.} 
Based on this, we define the following capacity limits for Berlin: (1) current maximal capacity (1,045 ICU beds) and (2) extended capacity by the end of April (2,267 ICU beds).

\subsubsection*{Lombardy (Italy)}
\begin{figure}
\centering
\includegraphics[width=0.48\linewidth]{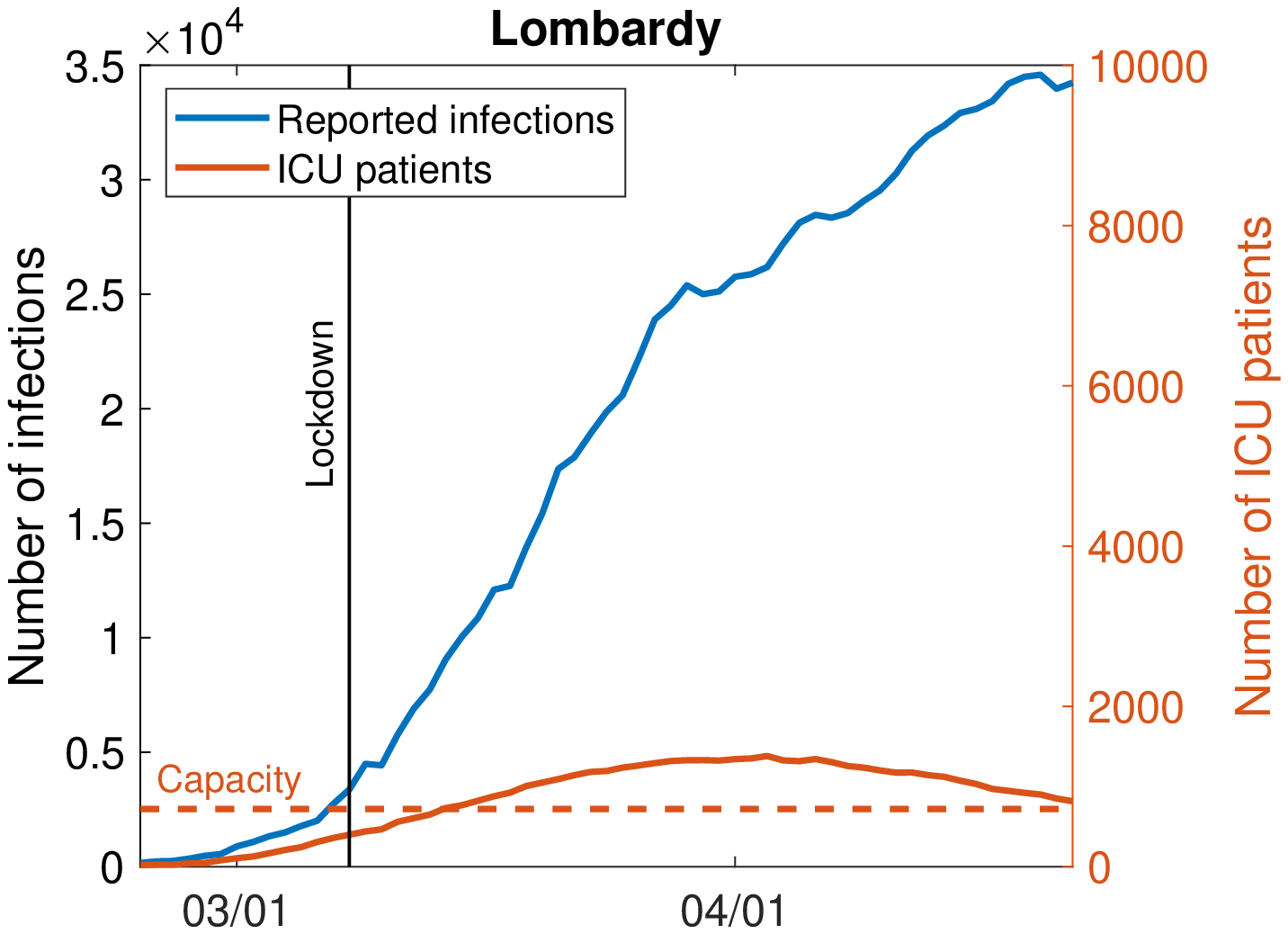}
\includegraphics[width=0.49\linewidth]{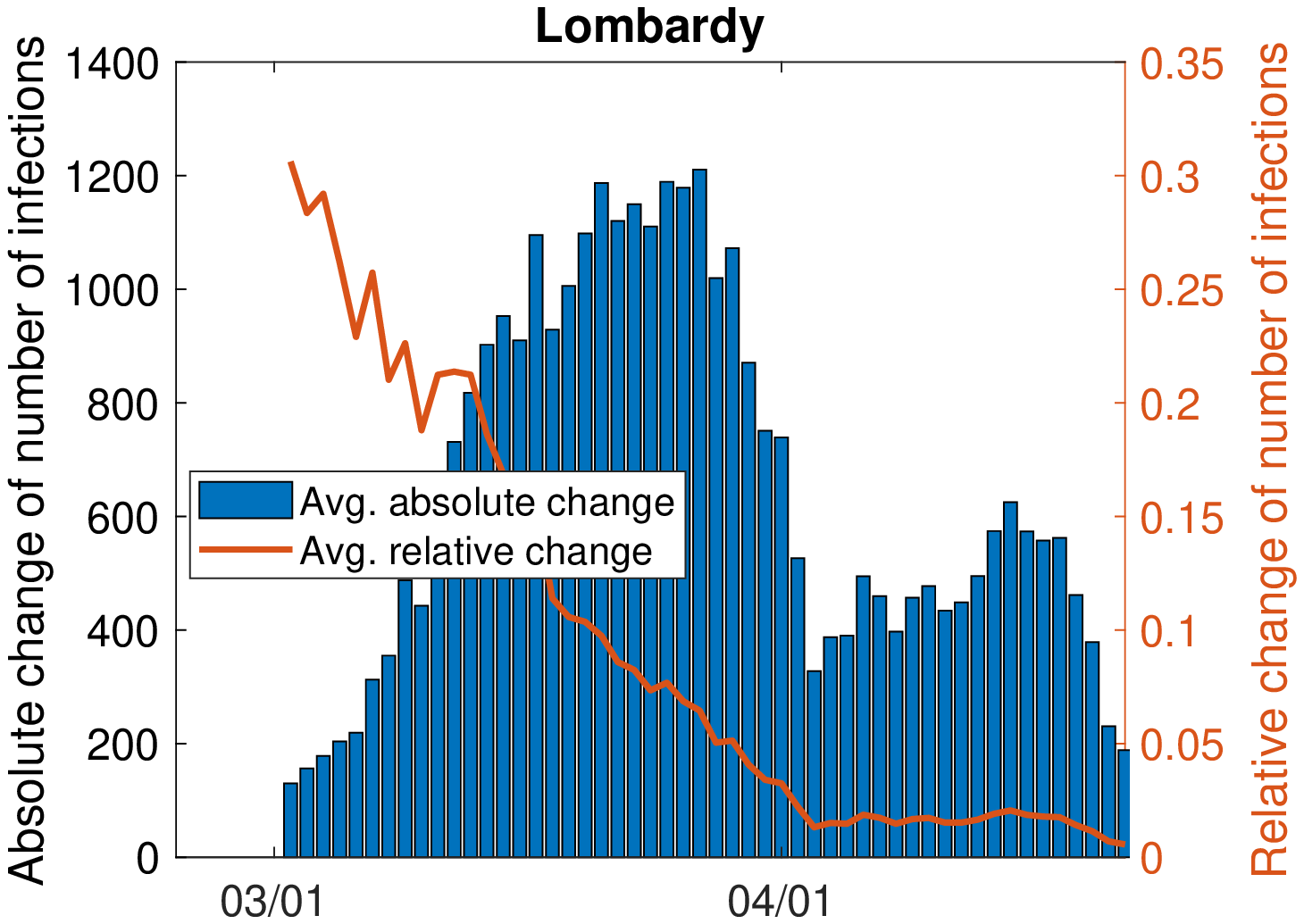}
\caption{Number of reported infections and intensive care patients (left) and absolute and relative daily change of the number of infections (right, averaged over the last 7 days, since 100 infections) in Lombardy until April 21, 2020 (based on data from the \textit{Presidenza del Consiglio dei Ministri - Dipartimento della Protezione Civile})}
\label{Figure:Data_Lombardy}
\end{figure}

Lombardy is one of the 20 administrative divisions of Italy and has about 10 million inhabitants.\footnote{\url{http://demo.istat.it/bil2018/index04.html}}
For each day between February 24, 2020, and April 21, 2020, we retrieved the number of reported COVID-19 patients and the number of COVID-19 patients in ICUs for Lombardy from the \textit{Presidenza del Consiglio dei Ministri -- Dipartimento della Protezione Civile} (Figure \ref{Figure:Data_Lombardy}, left).\footnote{\url{https://github.com/pcm-dpc/covid-19/blob/master/schede-riepilogative/regioni/dpc-covid19-ita-scheda-regioni-20200421.pdf}}
The total number of COVID-19 infections increased from 166 on February 24, 2020, to 33,978 on April 21, 2020. The number of ICU patients increased from 19 on February 24, 2020, to 1,381 on April 3, 2020. Since then, it decreased to 851 on April 21, 2020. Figure \ref{Figure:Data_Lombardy} (right) depicts the daily (absolute and relative) change in the number of reported infections over the previous 7 days.
On March 8, 2020, Italian Prime Minister Giuseppe Conte put the whole Lombardy under quarantine.\footnote{\url{https://www.nytimes.com/2020/03/07/world/europe/coronavirus-italy.html}} 
Assuming that at least 7 days are needed before the quarantine makes an impact, we restrict our analysis to data from February 24, 2020 to March 15, 2020. The overall daily growth rate for this period (with a minimum of 100 cases) is 22\% ($\sqrt[21]{10{,}043/166}=1.22$) and it decreased to 17\% in the week between March 8, 2020, and March 15, 2020 ($\sqrt[7]{10{,}043/3,372}=1.17$). 


Before the SARS-CoV-2 pandemic, the ICU capacities in Lombardy were estimated to be 720 beds \citep{Grasselli2020}, however, some hospitals have increased their number of ICU beds by the factor of 4--5 during the pandemic.\footnote{\url{https://www.esahq.org/esa-news/dynamics-of-icu-patients-and-deaths-in-italy-and-lombardy-due-to-covid-19-analysis-updated-to-30-march-day-38-evening/}}

\subsubsection*{Madrid (Spain)}

\begin{figure}
\centering
\includegraphics[width=0.48\linewidth]{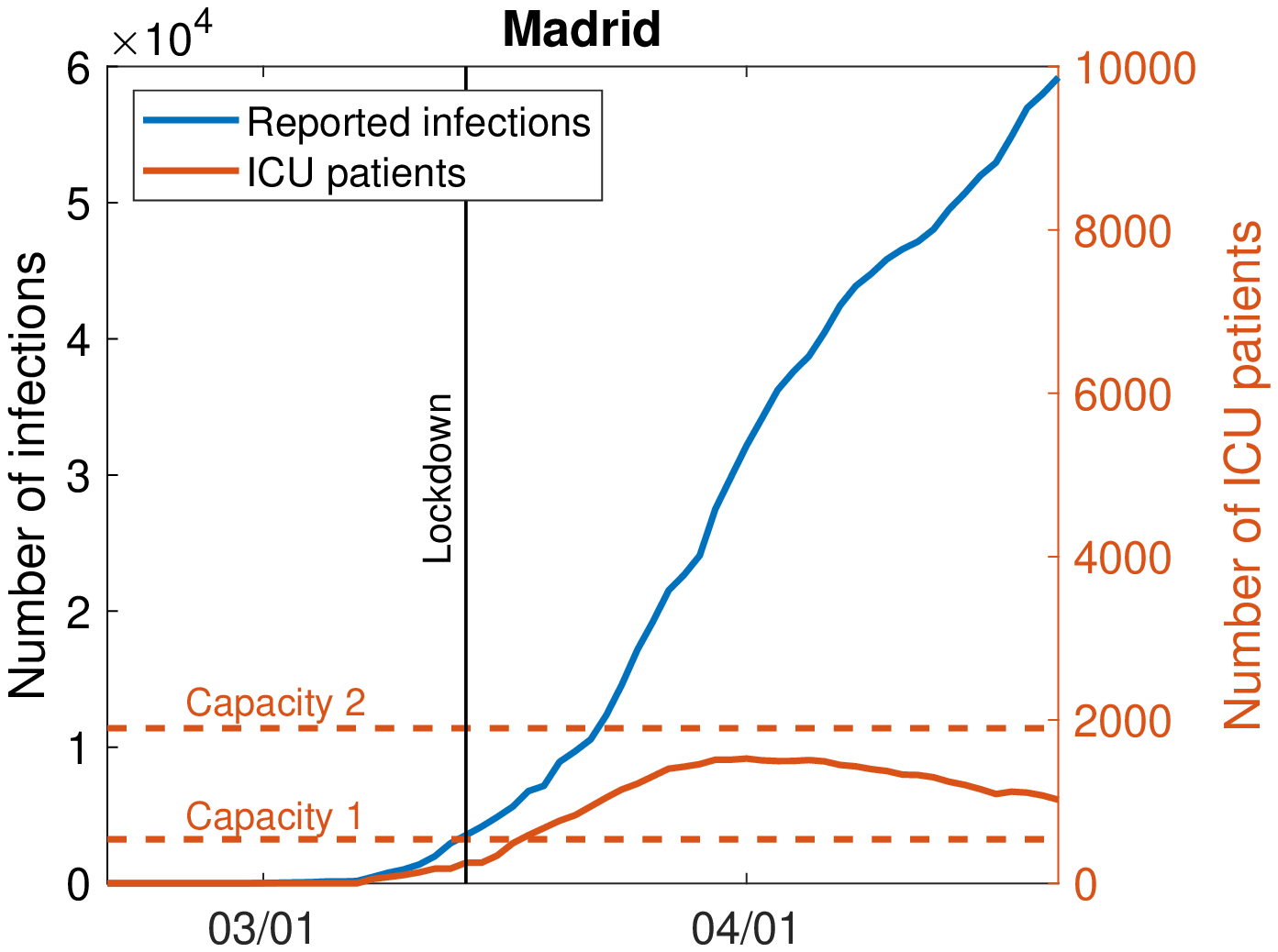}
\includegraphics[width=0.50\linewidth]{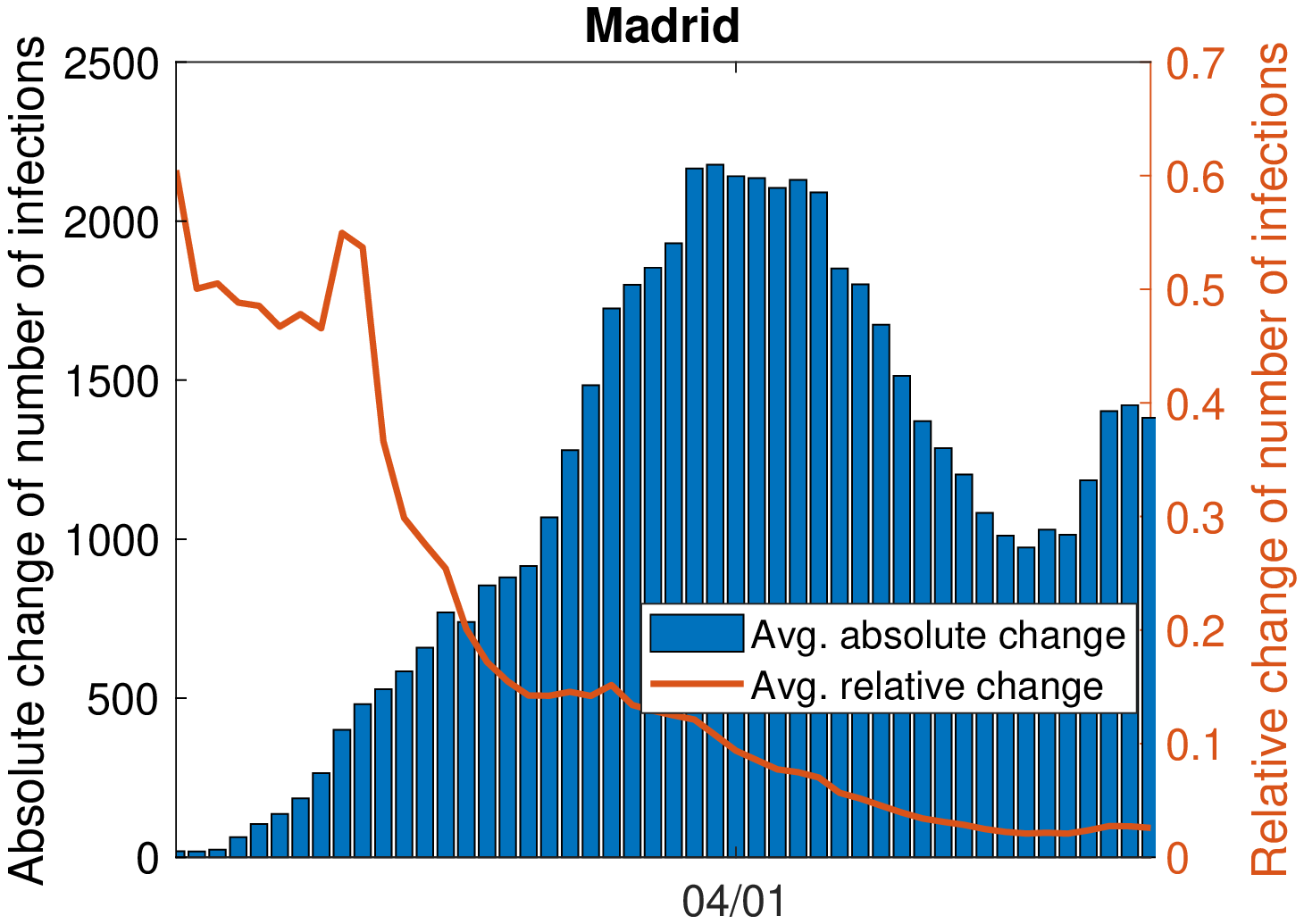}
\caption{Number of reported infections and intensive care patients (left) and absolute and relative daily change of the number of infections (right, averaged over the last 7 days, since 100 infections) in Madrid until April 21, 2020 (based on data from \textit{Datadista})}
\label{Figure:Data_Madrid}
\end{figure}

Madrid is one of the 17 autonomous communities in Spain and has about 
 6.6 million inhabitants.\footnote{\url{https://ec.europa.eu/growth/tools-databases/regional-innovation-monitor/base-profile/autonomous-community-madrid}}
For each day between February 25, 2020, and April 21, 2020, we retrieved the number of reported COVID-19 patients and the number of COVID-19 patients in ICUs for Lombardy from \textit{Datadista} (Figure \ref{Figure:Data_Madrid}, left).\footnote{\url{https://github.com/datadista/datasets/tree/master/COVID\%2019}}
The total number of COVID-19 infections increased from two on February 25, 2020, to 59,199 on April 21, 2020. The number of ICU patients increased from 53 on March 3, 2020, to 1,528 on April 1, 2020. Since then, it decreased to 1,024 on April 21, 2020. Figure \ref{Figure:Data_Madrid} (right) depicts the daily (absolute and relative) change in the number of reported infections over the previous 7 days.

On March 14, 2020, the Prime Minister of Spain Pedro Sánchez declared a nationwide State of Alarm and a national lockdown was imposed.\footnote{\url{https://administracion.gob.es/pag_Home/atencionCiudadana/Estado-de-alarma-crisis-sanitaria.html}}
Also for Madrid, we restrict our analysis to data until 7 days after the lockdown. Hence, our data set ranges from February 25, 2020, to March 21, 2020. The overall daily growth rate for this period (with a minimum of 100 cases) is 31\% ($\sqrt[16]{9{,}702/137}=1.31$) and it increased to 54\% in the week between March 7, 2020, and March 14, 2020 ($\sqrt[7]{3{,}544/175}=1.54$). 

An ICU capacity of 540 beds has been reported before the SARS-CoV-2 pandemic, during the pandemic this number has been tripled.\footnote{\url{https://elpais.com/espana/madrid/2020-04-14/madrid-calcula-que-300000-resident-es-han-sido-afectados-por-el-coronavirus.html}\\
\url{https://nbcmontana.com/news/nation-world/europe-faces-icu-crunch-rushes-to-build-field-hospitals}}

\subsection*{Results}
\begin{figure}
\centering
\includegraphics[width=0.47\linewidth]{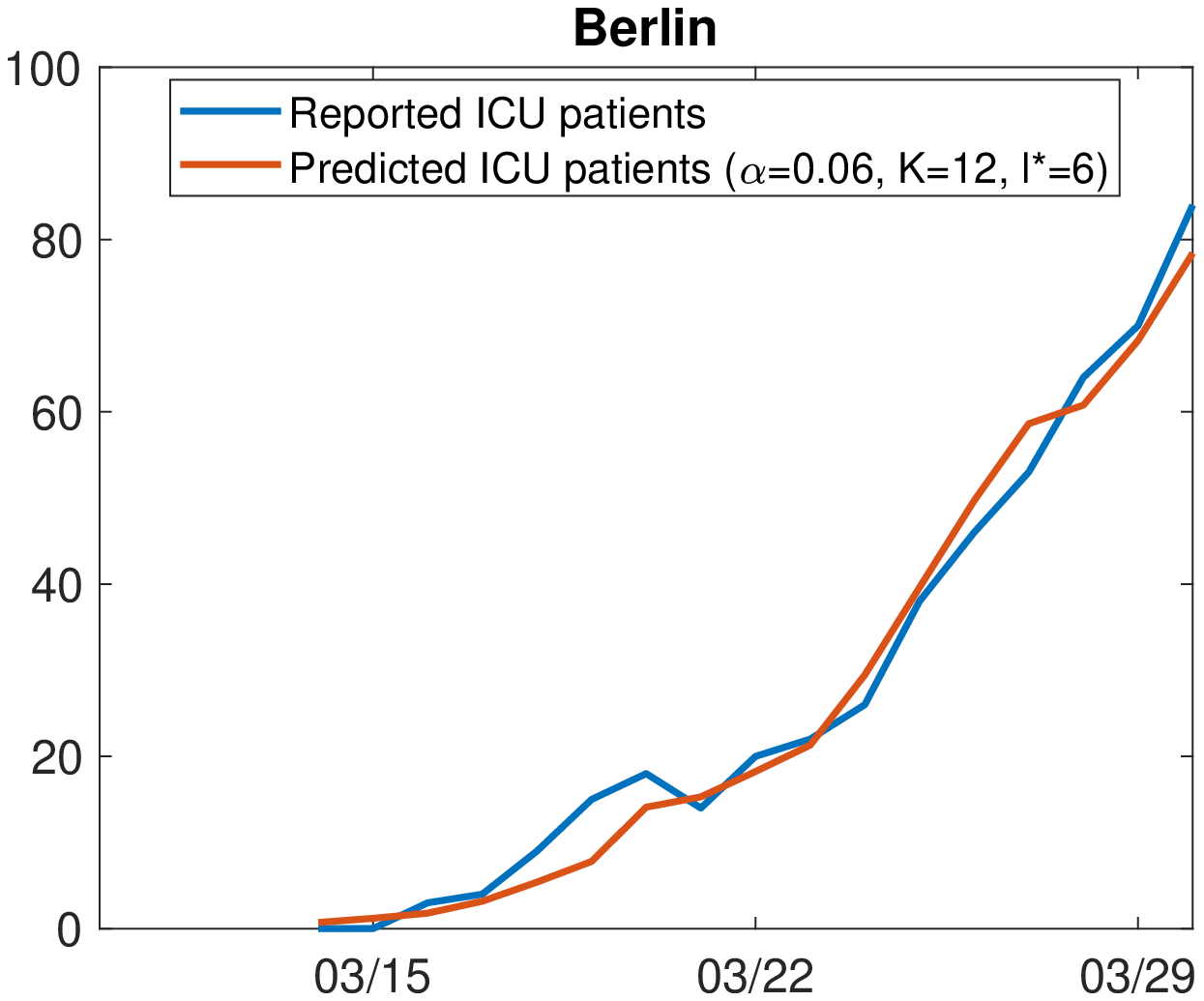}
\includegraphics[width=0.50\linewidth]{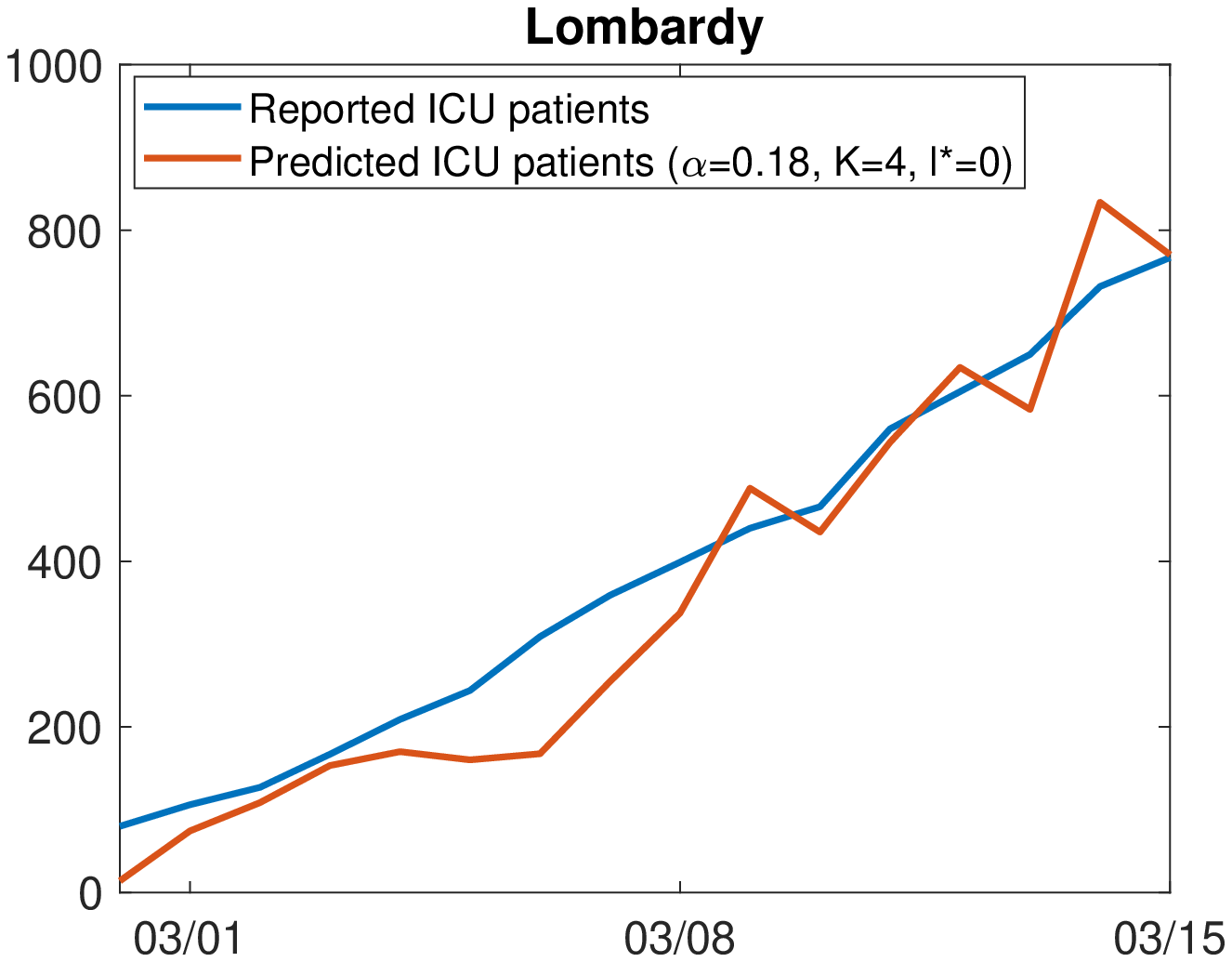}
\includegraphics[width=0.49\linewidth]{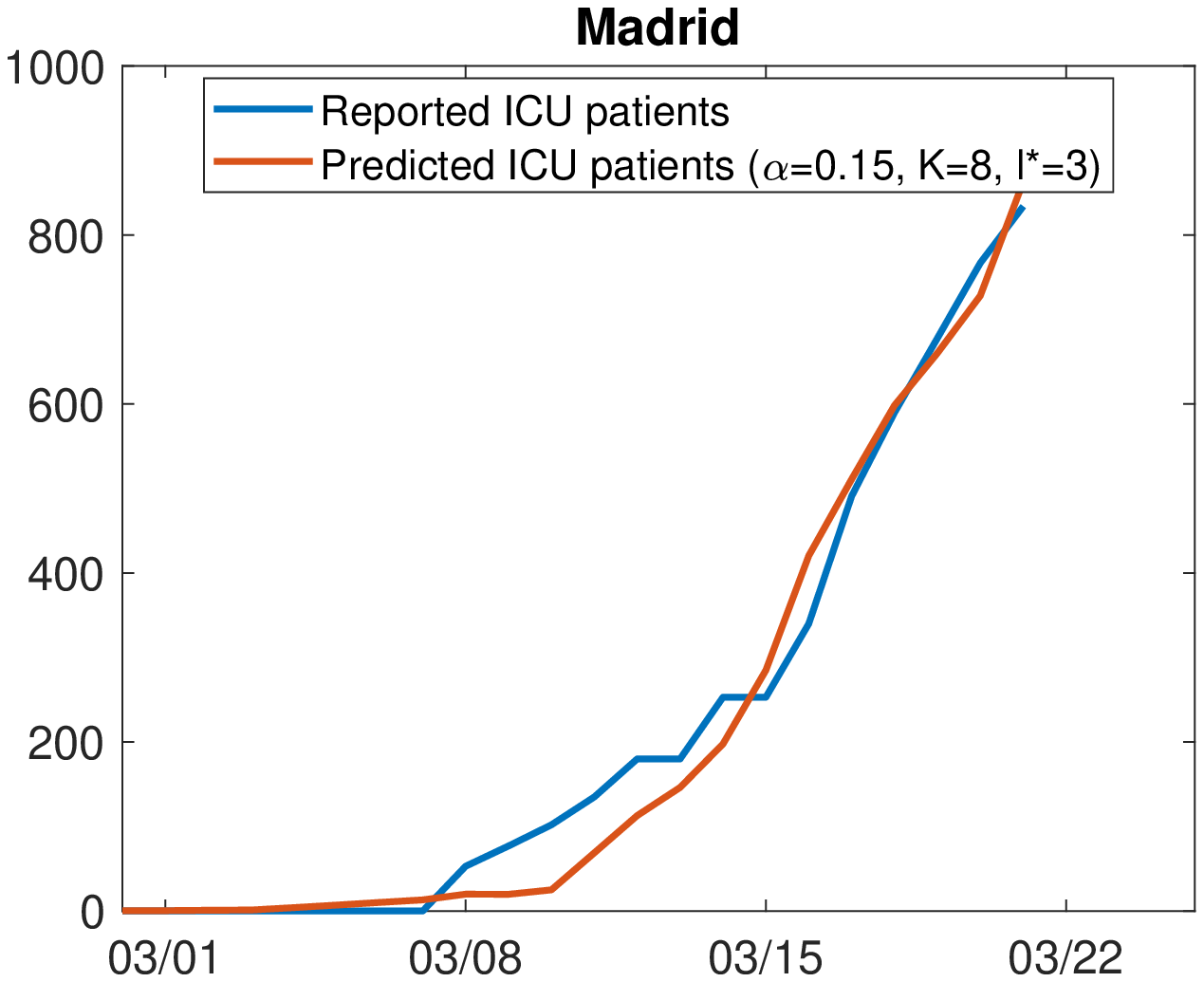}
\caption{Best model fit for the number of ICU patients with ICU rate $\alpha$, average stay in ICU $K$ and time lag $l^*$ between positive testing and entering ICU, separately for Berlin, Lombardy, and Madrid. }
\label{Figure:ModelFit_Berlin}
\end{figure}

First, we estimated the parameters of the model separately for Berlin, Lombardy, and Madrid (see Table \ref{tab:my_label}). 
For Berlin, the best fit was found for an ICU rate $\alpha$ of 6\%, a time lag of 6 days, and an average stay of 12 days in ICU. For Lombardy and Madrid, in contrast, a better fit was found for larger ICU rates (18\% and 15\%), shorter ICU stays (4 and 8 days), and a shorter time lag (0 and 3 days). The RMSE for Berlin is considerably lower than for Lombardy and Madrid, which can be explained by the lower number of ICU patients. The squared correlation coefficient lies between 0.96 and 0.98 for all three regions. 
In Figure \ref{Figure:ModelFit_Berlin}, we compare the reported ICU patients with the predicted ICU patients based on the estimated parameters. 

\begin{table}
    \centering
    \begin{tabular}{lrrrrrr}\toprule
         Region &$T$&  $\alpha$ & $K$ & $l^*$ & RMSE & $\rho^2$  \\\midrule
         Berlin &30& 0.06&   12&    6&    3.37&    0.98\\
         Lombardy &21& 0.18&  4 & 0  & 65.29   & 0.96 \\
         Madrid & 31&0.15  &  8 &3&   39.50&   0.98\\\bottomrule
    \end{tabular}
    \caption{Parameters for the best model fit separately for Berlin, Lombardy, and Madrid. $T$, number of days for which the model was fitted; $\alpha$, ICU rate; $K$, average stay in ICU; $l^*$, time lag between positive testing and entering the ICU; RMSE, root mean squared error; $\rho^2$, squared correlation coefficient.}
   \label{tab:my_label}
\end{table}

\begin{figure}
\centering
\includegraphics[width=0.49\linewidth]{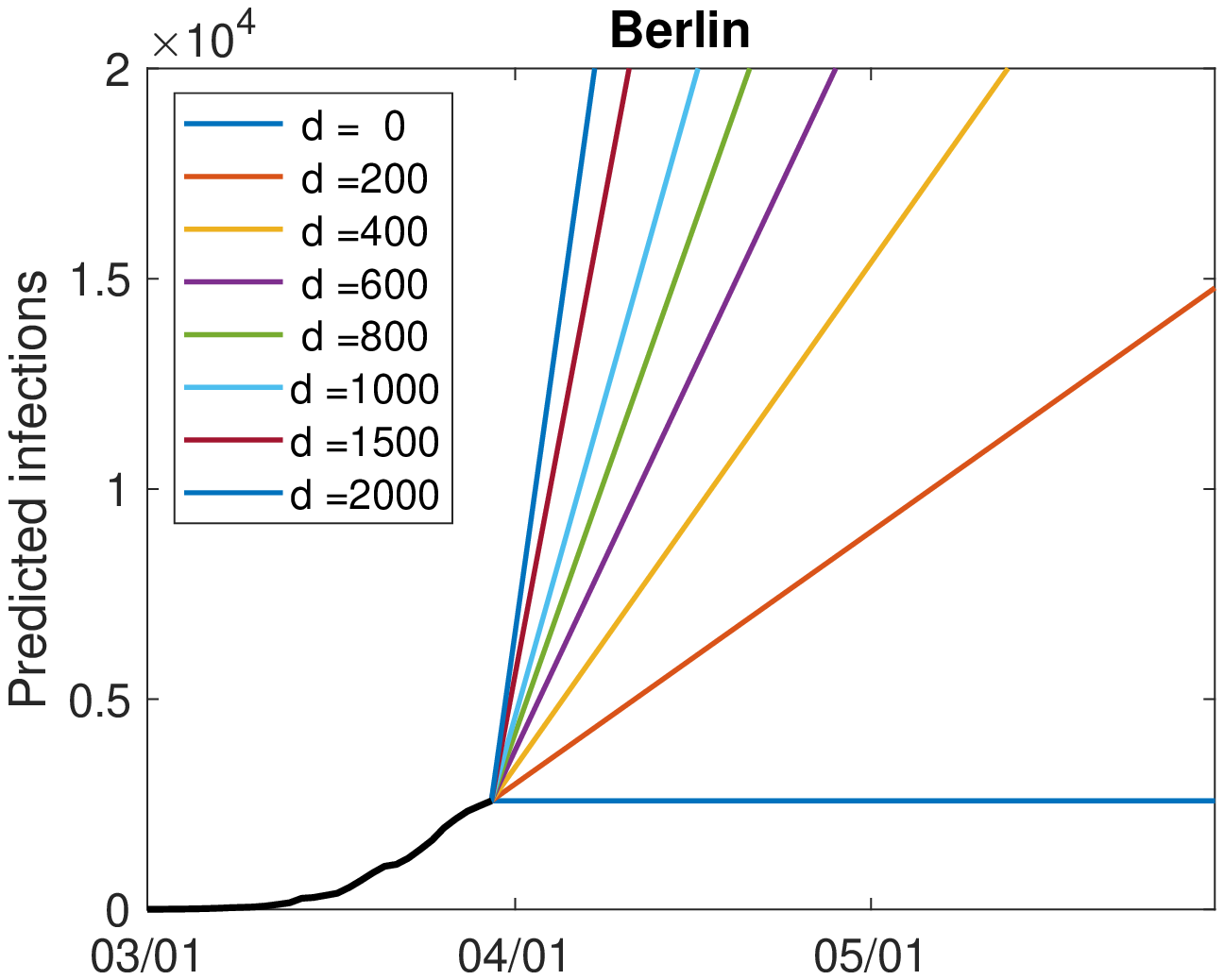}
\includegraphics[width=0.49\linewidth]{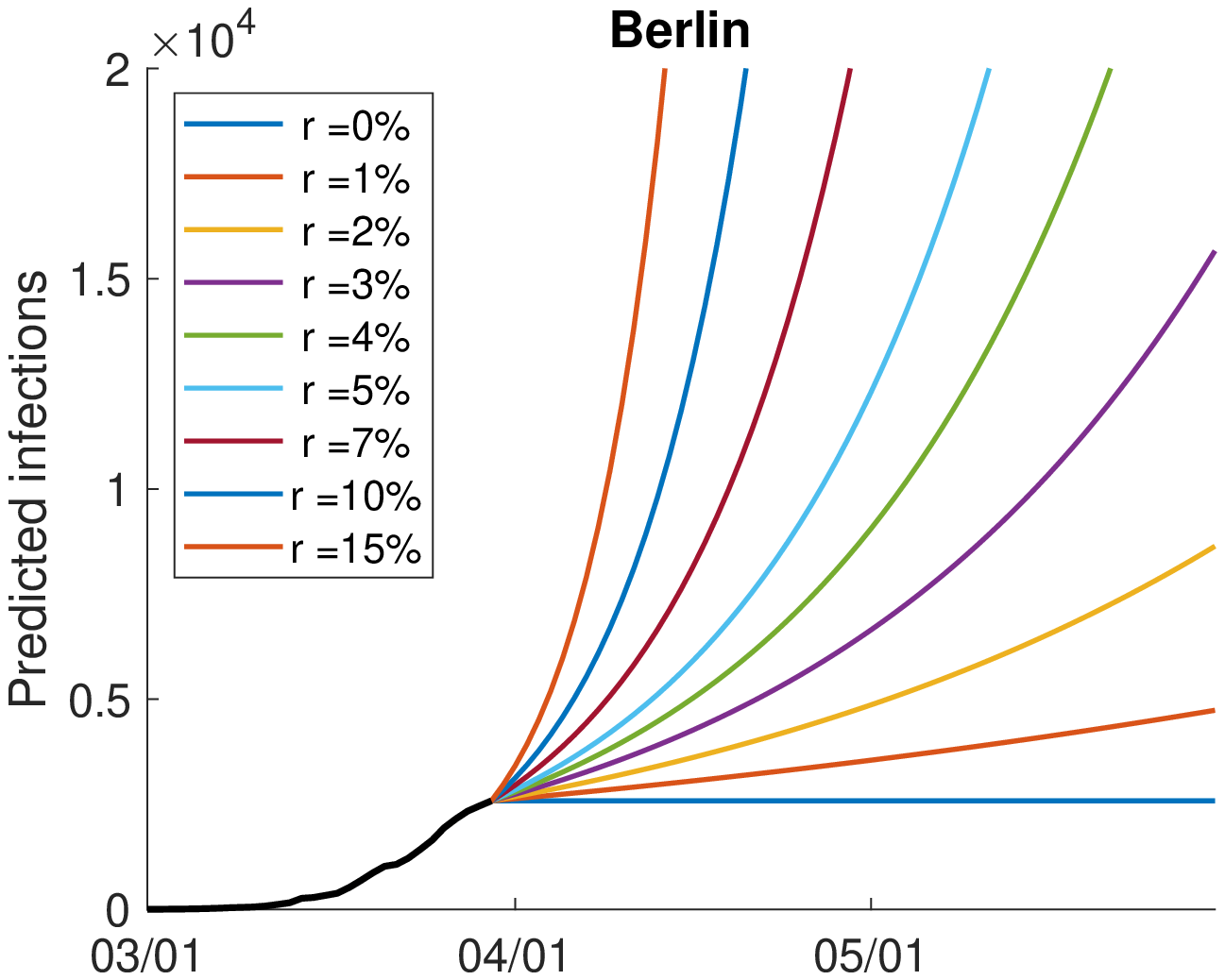}\\
\vspace{0.3cm}
\includegraphics[width=0.49\linewidth]{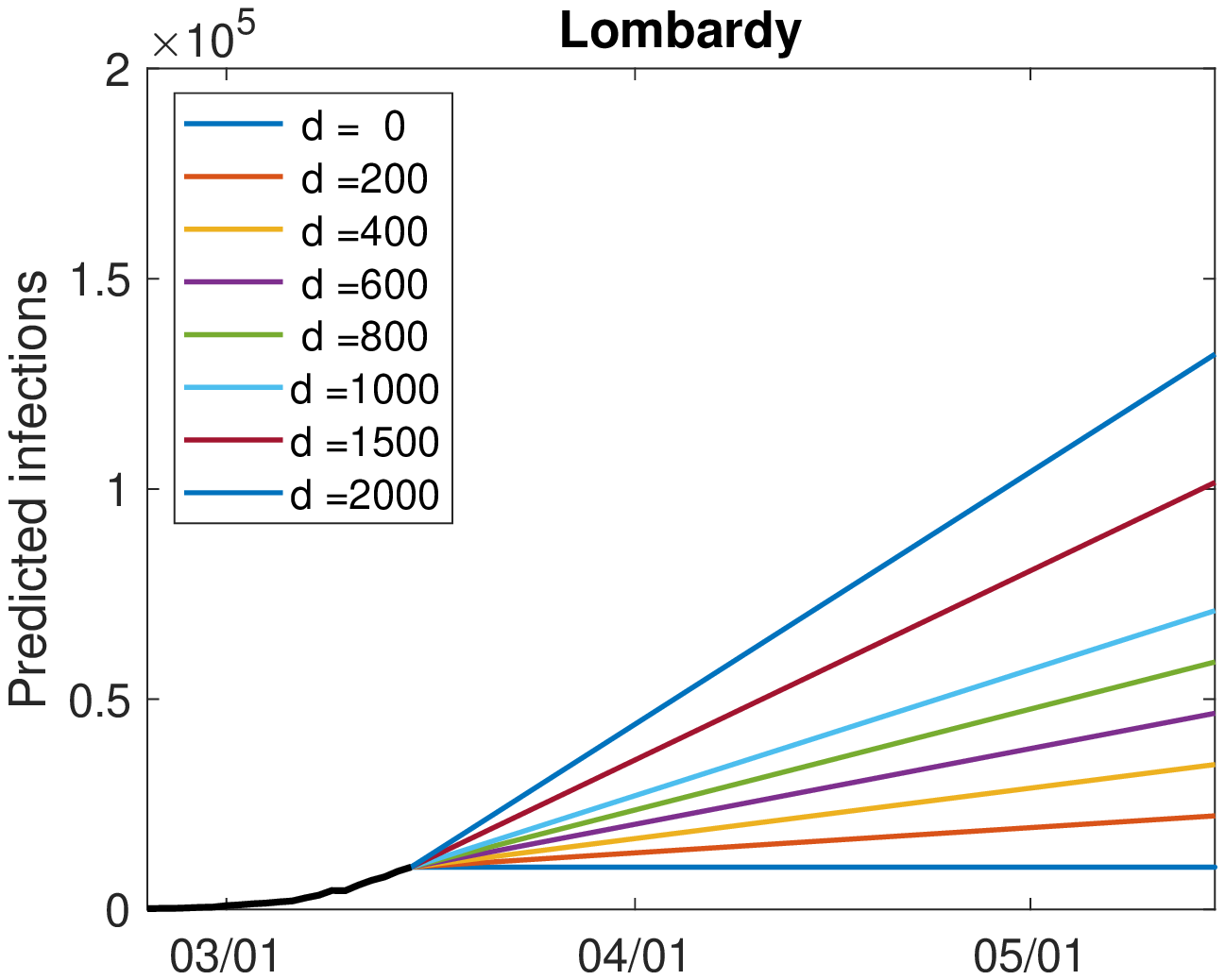}
\includegraphics[width=0.49\linewidth]{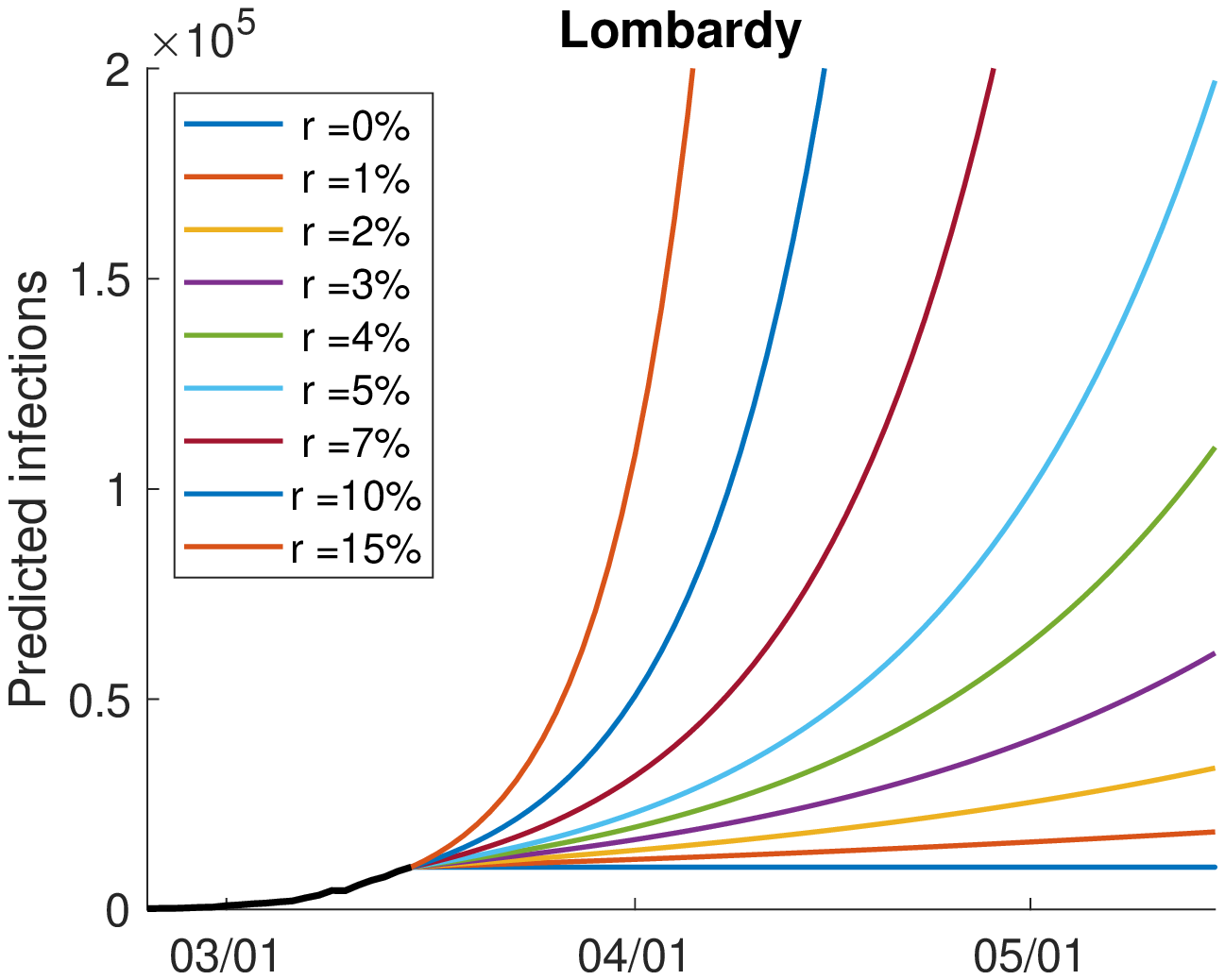}\\
\vspace{0.3cm}
\includegraphics[width=0.49\linewidth]{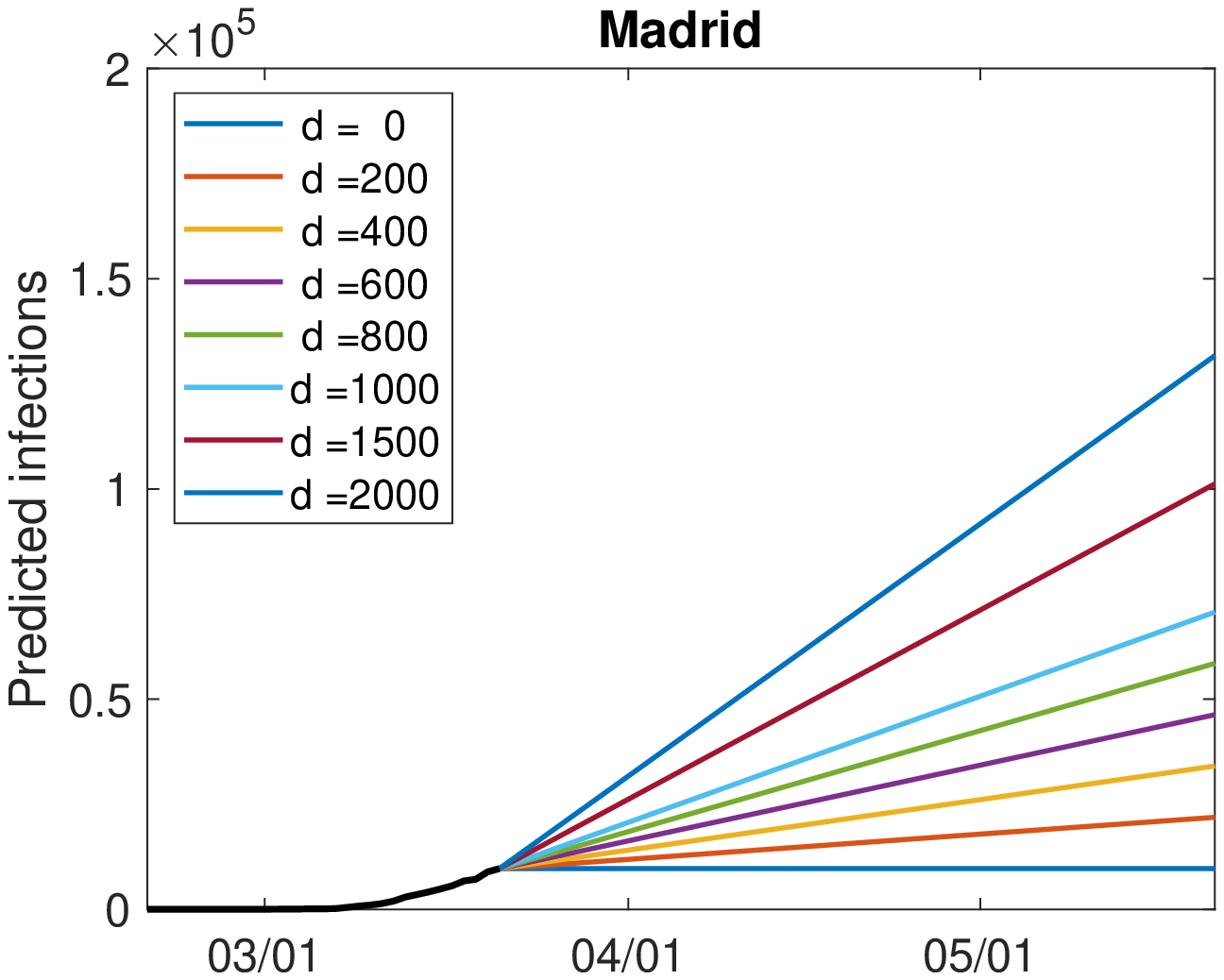}
\includegraphics[width=0.49\linewidth]{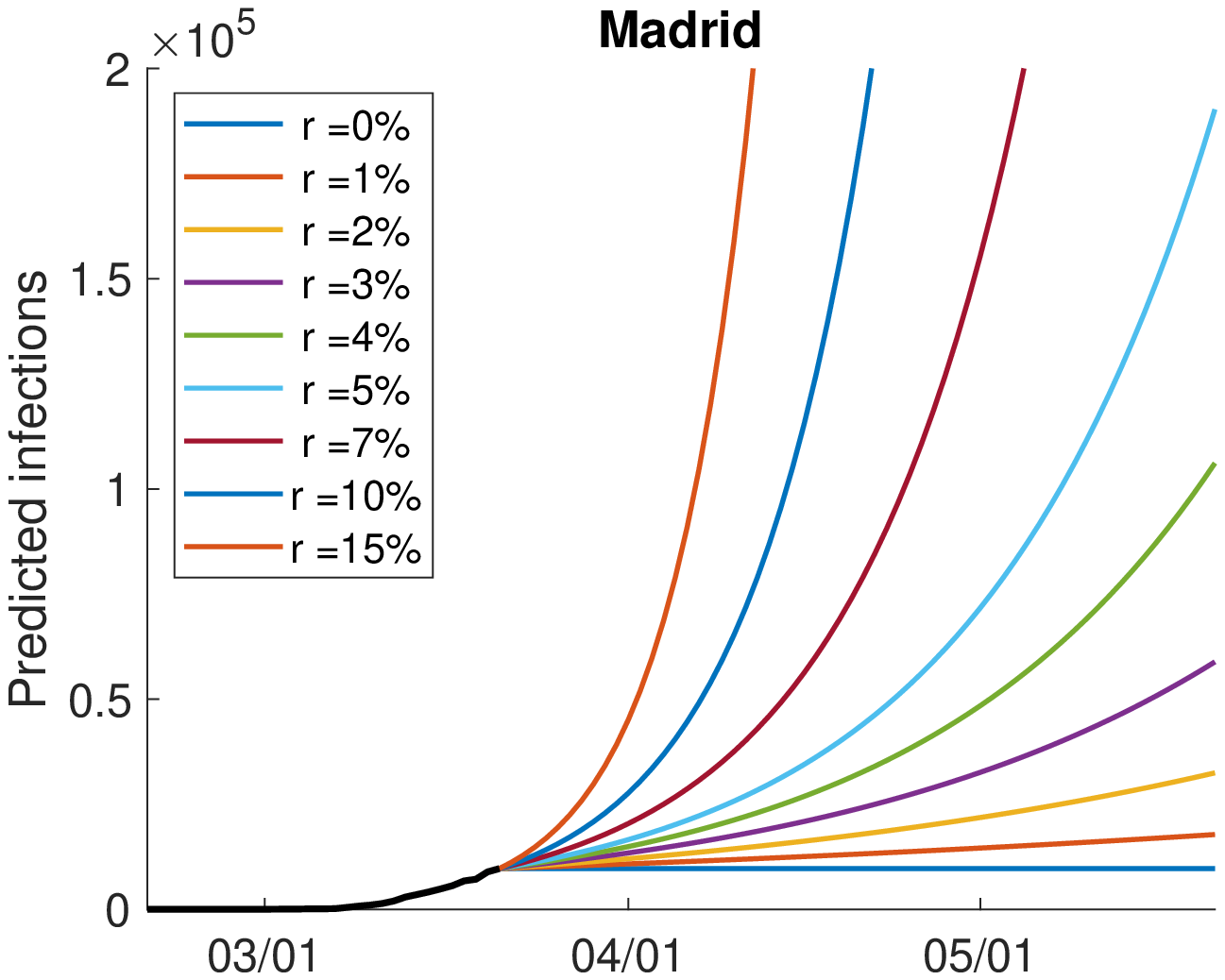}
\caption{Predicted number of infections assuming different linear (left) and exponential growth (right), separately for Berlin, Lombardy, and Madrid. The black line corresponds to the true observations until seven days after the lockdown.}
\label{Figure:PredictedInfections}
\end{figure}

\begin{figure}[h!]
\centering
\includegraphics[width=0.49\linewidth]{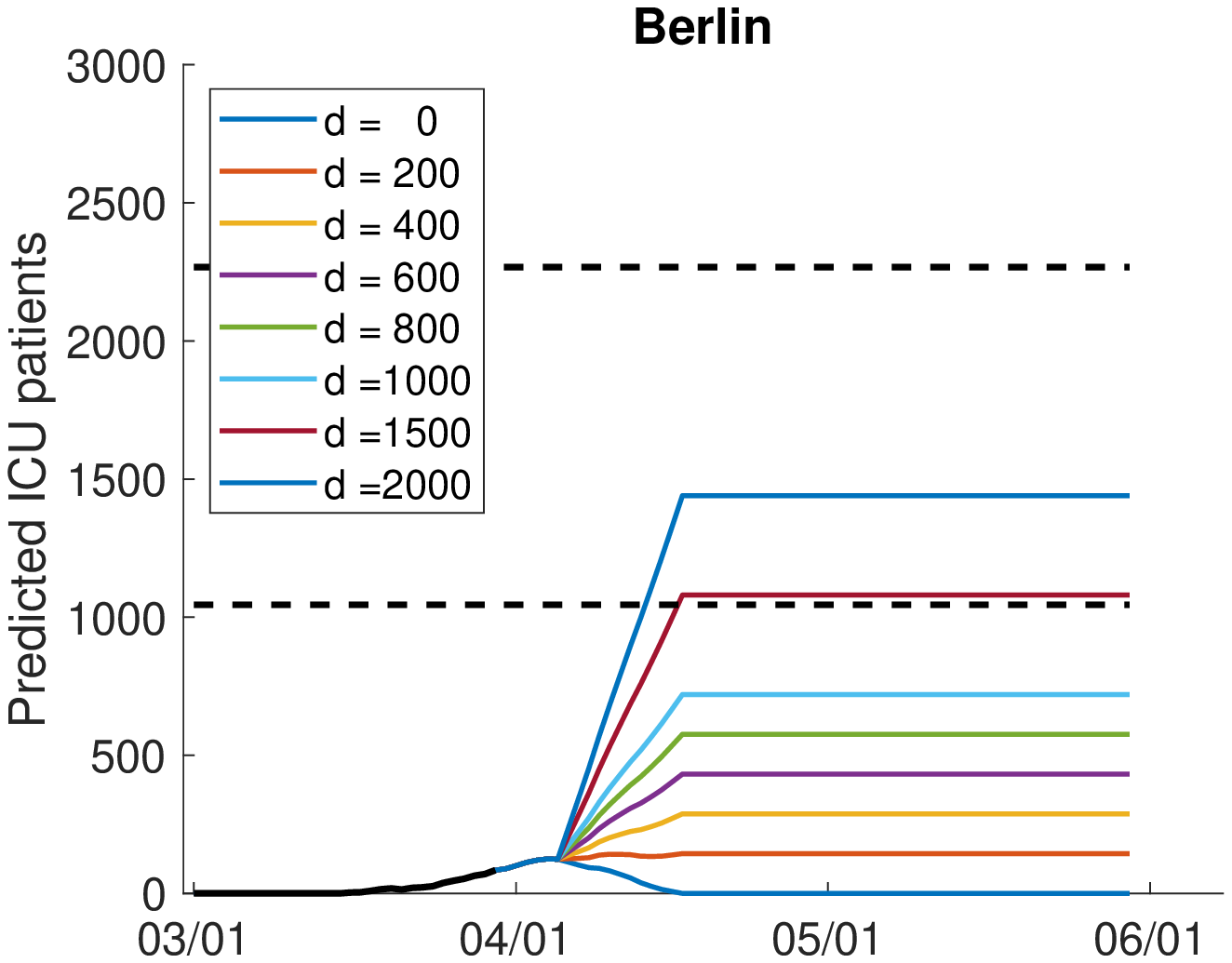}
\includegraphics[width=0.49\linewidth]{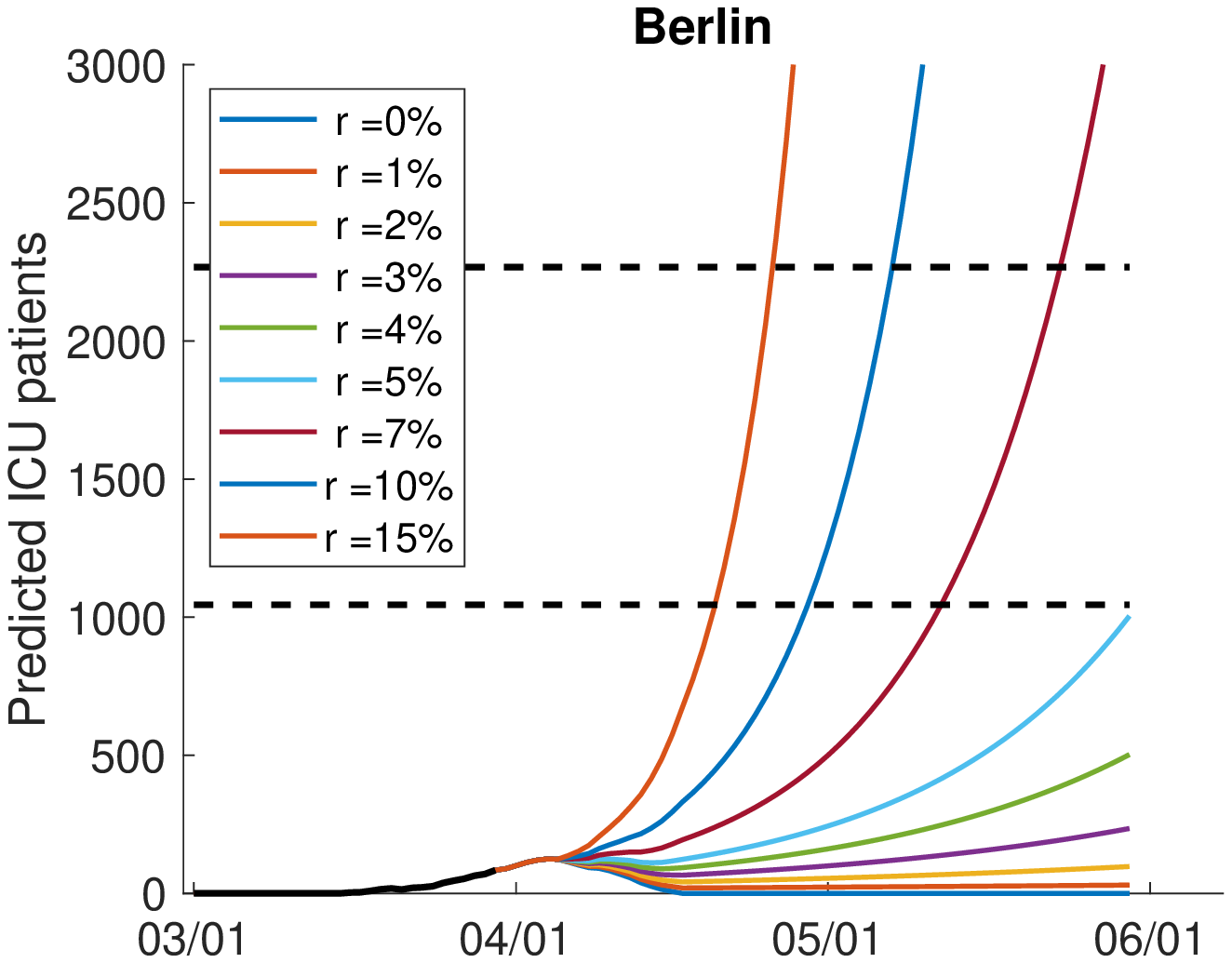}\\
\vspace{0.3cm}
\includegraphics[width=0.49\linewidth]{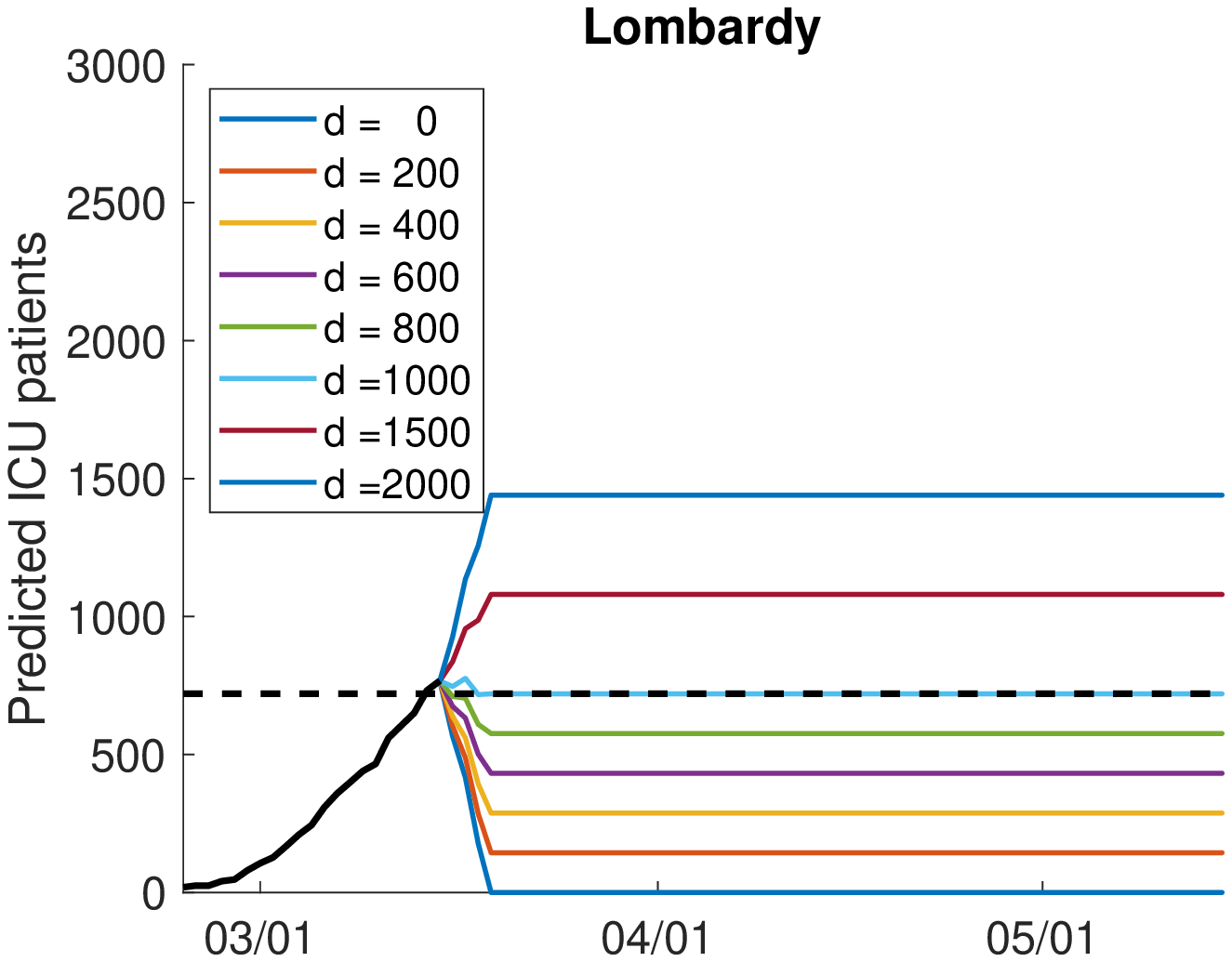}
\includegraphics[width=0.49\linewidth]{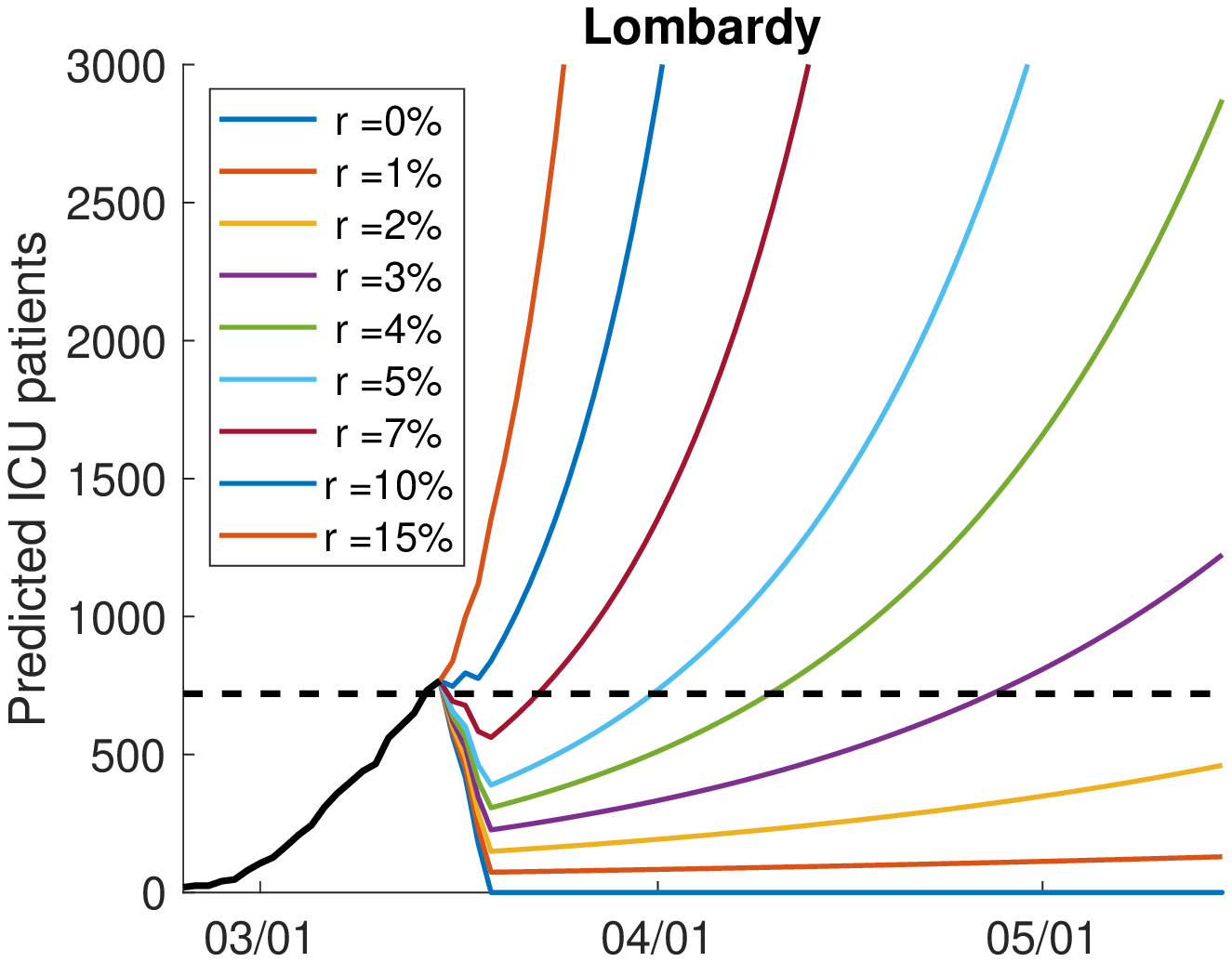}\\
\vspace{0.3cm}
\includegraphics[width=0.49\linewidth]{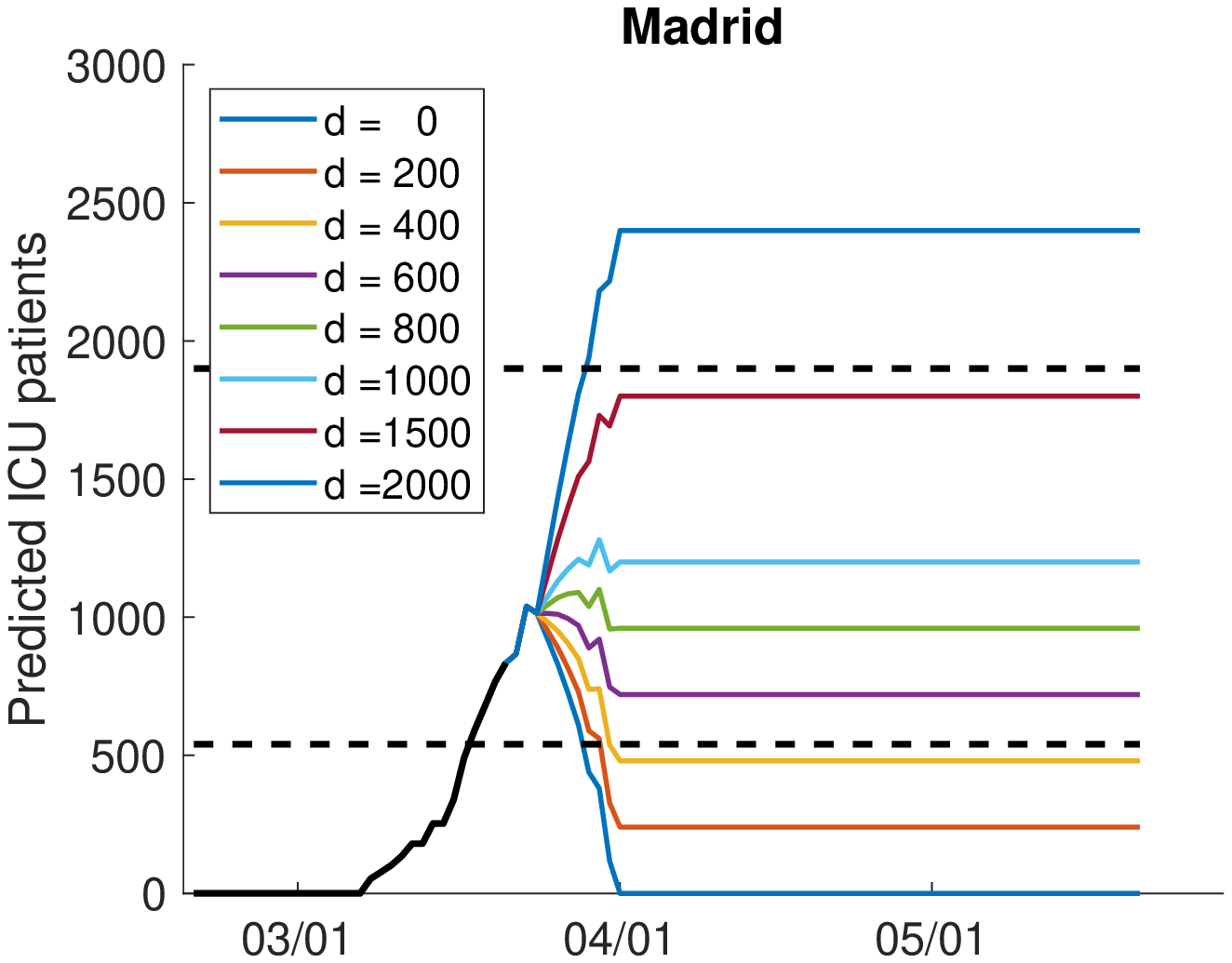}
\includegraphics[width=0.49\linewidth]{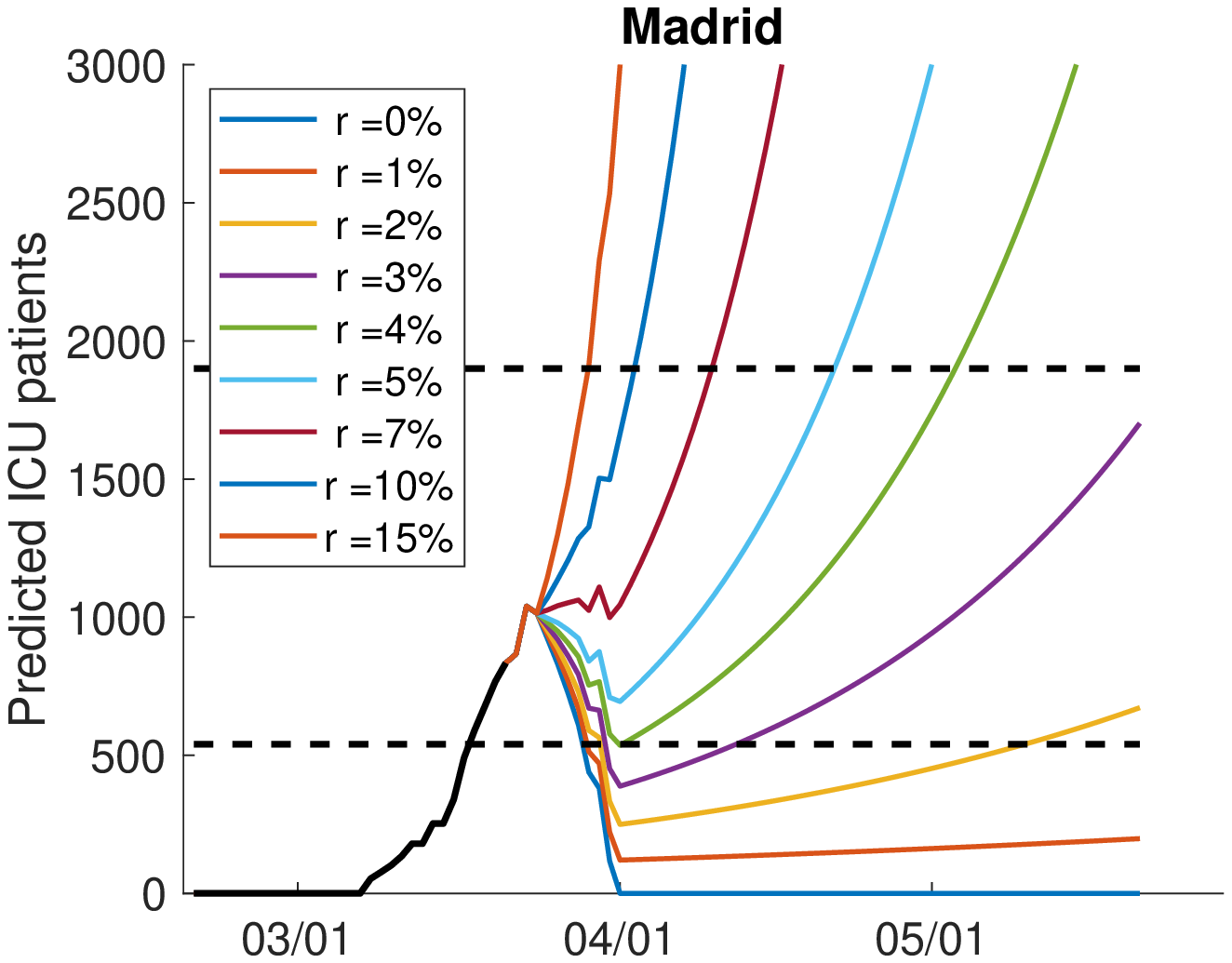}
\caption{Predicted number of ICU patients assuming different linear (left) and exponential growth (right), separately for Berlin, Lombardy and Madrid. The black line corresponds to the true observations until seven days after the lockdown. The dashed black line indicates the estimates of capacity.}
\label{Figure:PredictedICUPatients}
\end{figure}

Second, based on the models with the estimated parameters, we now predict the future development of (1) the number of infections (see Figure \ref{Figure:PredictedInfections}) and (2) the number of ICU patients  (see Figure \ref{Figure:PredictedICUPatients}).
Here, we will make a simplified assumption that the growth can be approximated as either linear or exponential. We show the sensitivity of the results for linear growth with the slope $d$ between $0$ and $2{,}000$ as well as exponential growth rates between 0\% and 15\%. A growth rate of 0\% means that from one day to another, no more new infections are reported, which is an unrealistic extreme case. The other extreme, 15\%, corresponds to a duplication of reported infections every 5 days. 
Please note that the daily growth rate in the last week before the strong containment measures came into force was 10\% for Berlin, 17\% for Lombardy, and 54\% for Madrid. 

In Figure \ref{Figure:PredictedICUPatients}, the predicted number of ICU patients is additionally compared with the capacity limits introduced in the data section. 
A linear growth in the number of infections leads in the long-run to a constant number of ICU patients, which corresponds to $\alpha \cdot d \cdot K$. This means, a share of $\alpha$ of the daily new infections $d$ needs intensive care for $K$ days.
Regarding exponential growth, our estimated parameters show that for Berlin only growth rates between 0\% and 4\% guarantee that the number of patients at the end of May remains below the current maximal capacity, a rate of up to 5\% guarantees to stay below the extended capacity by the end of May (but exceeds the capacity in June). For Lombardy and Madrid, it can be seen that the capacity was already exceeded before the containment measures came into force.

\begin{figure}
\centering
\includegraphics[width=0.32\linewidth]{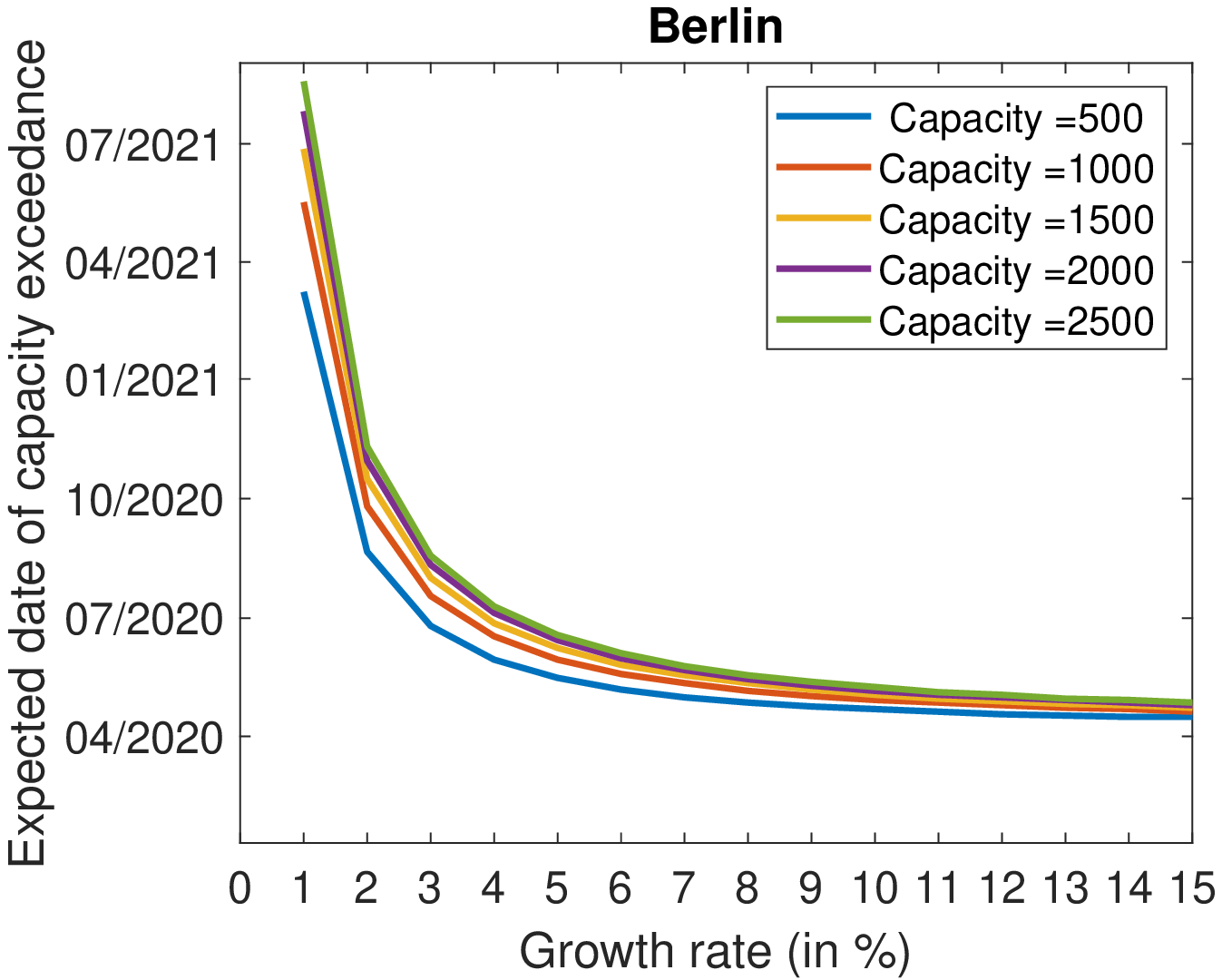}
\includegraphics[width=0.32\linewidth]{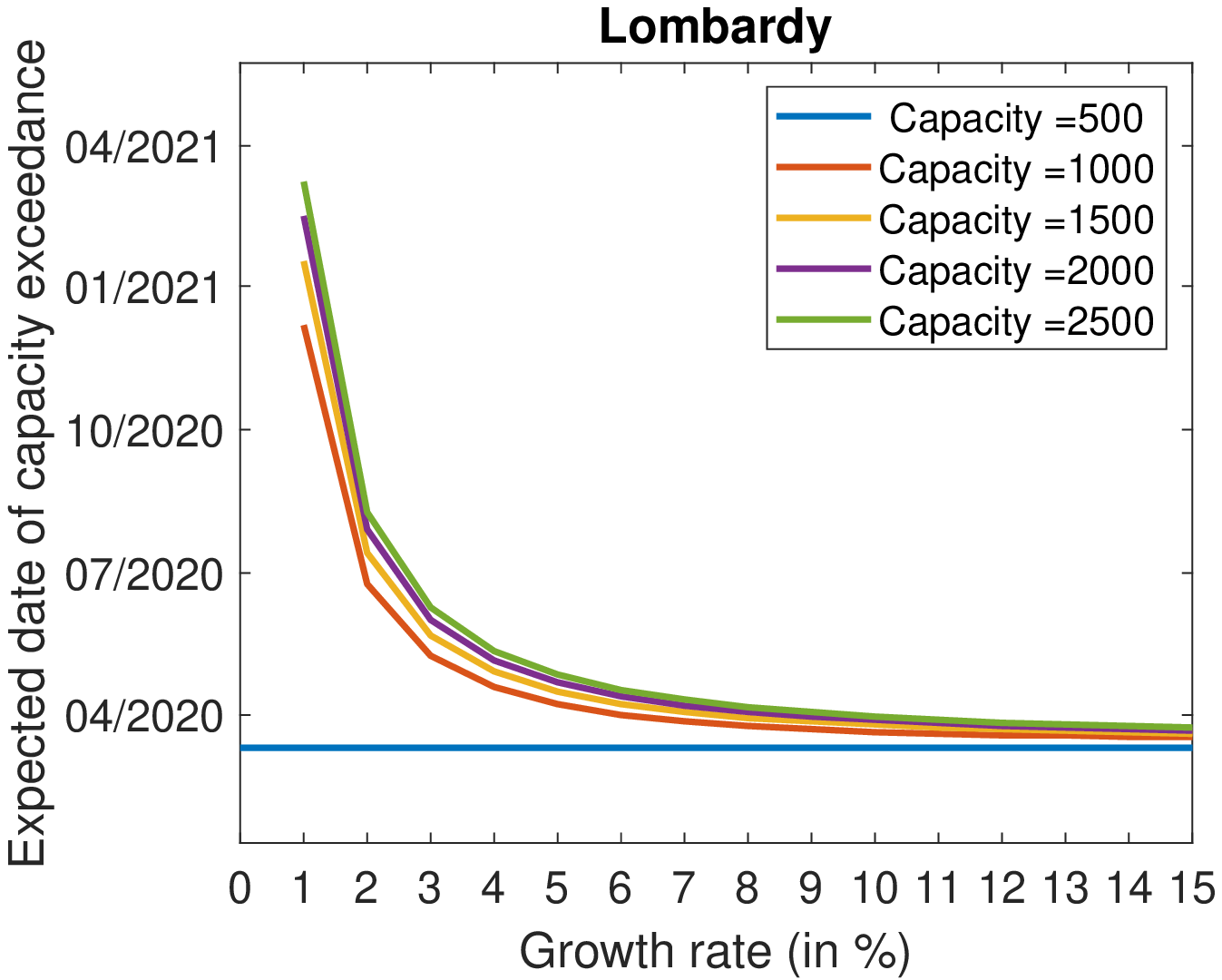}
\includegraphics[width=0.32\linewidth]{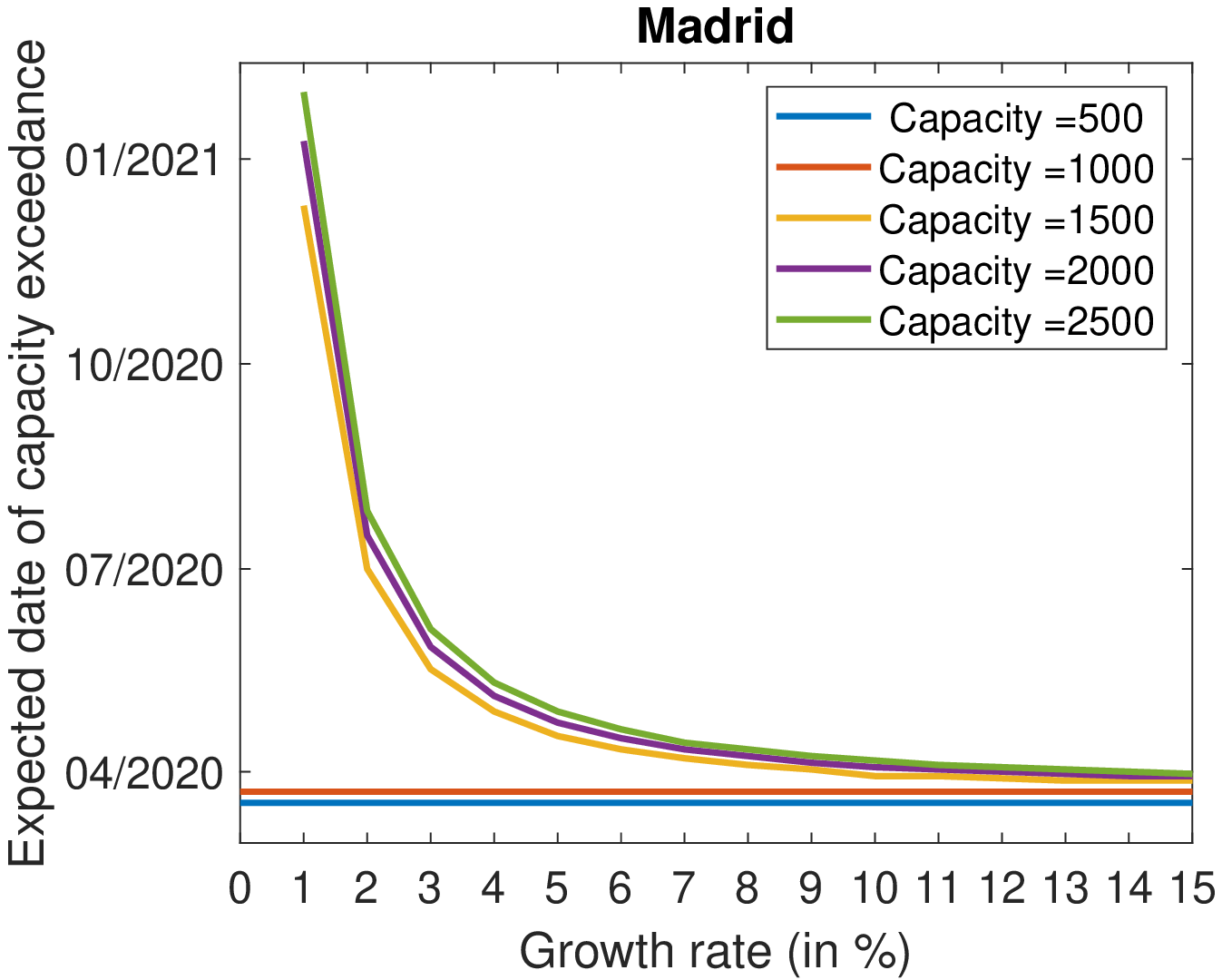}
\caption{Expected dates of capacity exceedance for different growth rates and exemplary capacities, separately for Berlin, Lombardy and Madrid.}
\label{Figure:CapacityExceedance}
\end{figure}

Figure \ref{Figure:CapacityExceedance} depicts the relation between different growth rates (0\%--15\%) and the dates of capacity exceedance for exemplary capacities of 500, 1,000, 1,500, 2,000, and 2,500 for Berlin, Lombardy, and Madrid. 
It demonstrates how the approach can be applied also for different phases of exponential growth or capacity extensions.

\section*{Discussion}

The ongoing COVID-19 pandemic deeply concerns policymakers and health personnel to take the right actions to slow down the spreading of SARS-CoV-2. In several regions including Lombardy and Madrid, the need for ICUs surpassed the available capacity and not all COVID-19 patients who needed intensive care could be treated \citep{Grasselli2020}. To avoid a triage, several countries, including Germany, Italy, and Spain, decided to apply containment measures -- ranging from temporary closures of schools and kindergartens to travel restrictions, compulsory mask wearing, and curfew. Those measures, however, disrupt economic activity and lead to the risk of a recession. Thus, policymakers are confronted with the difficulty of making decisions that severely affect the healthcare system, the global economy as well as the everyday life of many individuals while facing a large amount of uncertainty within the SARS-CoV-2 pandemic (e.g., the time until new treatments or vaccinations have been developed). Since one major bottleneck for the healthcare system is the ICU capacity, risk models are needed that allow policymakers to estimate the future ICU load to take appropriate measures.

So far, most forecast models for ICU load fit exponential growth (e.g., using ordinary least-squares in the log-space) to either the cumulated number of positive COVID-19 patients or directly to the number of ICU patients \citep{Deasy2020, Grasselli2020, Remuzzi2020}. For different regions in the UK, \citep{Deasy2020} used a Monte Carlo simulation to predict regional ICU capacities. Those models, however, assume that the initial exponential growth will hold over the forecast horizon (usually 14 days) and do not account for the alleviation of growth rates, e.g., due to containment measures. Moreover, these models do not exploit the underlying relationship between reported infections and ICU admissions, e.g., in terms of ICU rate. Here, we provide a simple, comprehensible, and transparent model that allows for predicting ICU load assuming linear or different exponential growth of the number of infections. 

We evaluated this model for Berlin, Lombardy, and Madrid, where we had access to the number of reported COVID-19 infections and ICU patients with COVID-19. Based on data before extensive containment measures made an impact, we first estimated the parameters ICU rate, average stay in ICU, and time lag. 
For Berlin, an ICU rate of 6\% was estimated which is close to an estimate of 5\% in \citep{SMC2020} and expectations of intensive care physicians in Berlin \citep{Bachetal2020}. \cite{Wu2020} also report an ICU rate of 5\% based on data for 72,312 cases in China. For Lombardy and Madrid, in contrast, a much higher ICU rate of 18\% and 15\%, respectively, resulted in the best model fit. This is in relative accordance with \cite{Grasselli2020} who reported an ICU rate of 16\% for Lombardy in the first two weeks of the SARS-CoV-2 pandemic. Please note here that the ICU rate does not only depend on age, but also on the total number of tests because more testing will lead to a higher number of mild cases and thus a lower ICU rate \citep{Ferguson2020}. Thus, the higher ICU rates found for Lombardy and Madrid might also reflect a more restrictive testing scenario in those areas \citep{Garcia2020,Gobierno2020}. In similar models, relying only on exponential growth of intensive care patients, ICU rates of 9--11\% have been assumed \citep{Remuzzi2020}. 

Regarding the average stay in ICU (until recovery or death), the best model fit resulted in 12 days for Berlin, 4 days for Lombardy, and 8 days for Madrid. Based on data from China, other forecast models assumed an average stay of 8 to 10 days \citep{Ferguson2020,Deasy2020}. For Lombardy, however, a median length of ICU stay of 9 days has been reported for the early phase \citep{Grasselli2020baseline}. Also for other regions in Italy and USA, longer ICU stays (about 15 days for ICU patients who remain alive and about 10--12 days for patients who die after ICU treatment) have been found \citep{Manca2020, Bhatraju2020}. 
In New York, COVID-19 patients typically need ICU care with ventilation for 11 to 21 days, some patients stay in ICU for 30 days \citep{Clukey2020}.

The time lag between positive testing and ICU admission was estimated to be 6 for Berlin, 0 for Lombardy, and 3 for Madrid. The time lag is known to be highly variable because COVID-19 patients may be tested at different time points due to different disease courses as well as regulatory and organizational issues (e.g., capacity of test units or eligibility to get a test). Thus, the low numbers for Lombardy and Madrid can be explained by the limited testing capacity in those regions, where tests are mainly administered for patients with severe COVID-19 symptoms \citep{Garcia2020,Gobierno2020}. Please note that the time lag does not reflect the time between the actual infection and ICU admission. 

We then used the models with the estimated parameters to predict the number of ICU patients for a fixed time horizon of two months. By assuming linear or exponential growth of different levels and different ICU capacities, we evaluated different scenarios and show the sensitivity towards the growth rate in the exponential phase of COVID-19. The further the expected dates of a capacity exceedance can be shifted into the future, the higher the likelihood that new treatments or a vaccination are available or that a larger share of the population has become immune against COVID-19, which would further lower the growth rate. Please note that our results only hold true in the early phase of the disease and need to be adapted in later phases when the growth deviates from exponential. 

To account for the dynamic situation of the deployment of ICU beds, we compare the predicted number of ICU patients to different estimates of ICU capacity (if available). For simplicity, we assumed that all those beds are available for only COVID-19 patients and thus did not account for different utilization rates which might lead to an earlier capacity exceedance in practice. 
Additionally, we did not account for shortage in highly-qualified medical personnel. 
Already prior to the COVID-19 associated increase in capacities, it had been a common phenomenon that ICU beds were unused for lack of qualified personnel. 
Please note that we here also did not differentiate between high and low care ICUs (i.e., with and without invasive ventilator) since we only had numbers on the total number of ICU patients with COVID-19. Future studies might also try to account for changing treatment strategies of COVID-19 patients. Guidelines have emerged \citep{Alhazzani2020}, but are likely
subject to change. Early reports on the outbreak emphasized the potential progression of COVID-19 pneumonia to acute respiratory distress syndrome (ARDS) \citep{Yang2020}, biasing towards early intubation and ventilation. However, a growing number of studies shows that a subgroup of patients seems to be clinically stable despite very low blood oxygenation levels and may benefit more from non-invasive oxygen supply and ventilation methods than early intubation \citep{Gattinoni2020}. Accumulating experience with COVID-19 patients will therefore result in shifting treatment strategies that can either increase or ease the strain on ICU capacities.

By testing the proposed model for three regions, we show that this approach is generally generalizable to data of other cities, states, or whole countries as long as the overall numbers of reported COVID-19 patients and the ICU capacities are available. When data on ICU patients with COVID-19 are additionally available, it can further be evaluated if our parameter estimates also hold true for other regions. An informative repository in Germany is the DIVI registry, in which most hospitals in Germany provide numbers of the current ICU load and capacity.\footnote{\url{https://www.intensivregister.de/}}

Although our predictions are based on a small data set, and involve unclear dynamics and a number of assumptions, we introduced here a simple statistical model that (1) produces reasonable estimates for ICU rate, length of ICU stay, and time lag and (2) accounts for different exponential growth rates as well as linear growth with different slopes.
Our analyses for Berlin, Lombardy, and Madrid demonstrate that a continued exponential growth rate led or would have led to an exceedance of the ICU capacity in the near future. The predicted capacity exceedance is also in line with other forecast models for the UK or the USA \citep{Deasy2020,IMHE2020}. 
Due to the containment measures in Berlin, Lombardy, and Madrid (e.g., closure of schools, travel restrictions, and curfew), the growth rate, however, has strongly decreased since then so that lower growth rates or linear growth seem to be most realistic. 
Since our proposed model is quite generic, we hope that this study will help others to estimate the future need of intensive care in addition to sophisticated epidemiological approaches such as compartment or transmission models \citep{Ferguson2006,Ferguson2020,Fox2020}. However, as already pointed out by others \citep[e.g.,][]{Deasy2020}, any forecast models in these early times of the current SARS-CoV-2 pandemic needs to be taken with caution and assumptions need to be updated when further data become available. In particular, future studies might model the number of administered tests as a confounding variable.

\bibliography{main}


\section*{Acknowledgements}

We are grateful to Prof.\,Dr.\,med.\,Michael M. Ritter for his clinical input and Hans-Ludwig Hackmack for revising the mathematical model.

\section*{Author contributions statement}
MR, JDH and KR designed the study. MR and KR developed the method. MR analyzed the data. MR, DO, FP, JDH and KR wrote the manuscript. 

\section*{Additional information}
The authors report no competing interests.





\end{document}